\documentclass[twocolumn]{aastex62}

\accepted{for publication in the Astrophysical Journal}

\newcommand{\teff}{T_{\rm eff}}
\renewcommand{\bv}{B\, -\, V}
\newcommand{\vi}{V\, -\, I_C}

\newcommand{\jk}{J\, -\, K_s}
\newcommand{\hk}{H\, -\, K_s}

\newcommand{\gr}{g\, -\, r}
\newcommand{\gi}{g\, -\, i}

\newcommand{\ebv}{E(B\, -\, V)}

\newcommand{\dmn}{(m\, -\, M)_0}
\newcommand{\dmnv}{(m\, -\, M)_V}

\shorttitle{Photometric Mass of Red Clump Giants}
\shortauthors{An et~al.}

\begin{document}

\title{Comparison of the Asteroseismic Mass Scale of Red Clump Giants with Photometric Mass Estimates}

\author{Deokkeun An}
\affiliation{Department of Science Education, Ewha Womans University, 52 Ewhayeodae-gil, Seodaemun-gu, Seoul 03760, Republic of Korea}

\author{Marc H.\ Pinsonneault}
\affiliation{Department of Astronomy, The Ohio State University, Columbus, OH 43210, USA}

\author{Donald M.\ Terndrup}
\affiliation{Department of Astronomy, The Ohio State University, Columbus, OH 43210, USA}

\author{Chul Chung}
\affiliation{Department of Astronomy and Center for Galaxy Evolution Research, Yonsei University, Seoul 03722, Republic of Korea}

\correspondingauthor{Deokkeun An}
\email{deokkeun@ewha.ac.kr}

\begin{abstract}

Asteroseismology can provide joint constraints on masses and radii of individual stars. While this approach has been extensively tested for red giant branch (RGB) stars, it has been more difficult to test for helium core-burning red clump (RC) giants because of the lack of fundamental calibrators. To provide independent mass estimates, we utilize a number of widely used horizontal-branch (HB) models in the literature and derive photometric masses from a comparison with $griBVI_CJHK_s$ photometry. Our selected models disagree with each other on the predicted mass-luminosity-temperature relation. We adopt first-order corrections on colors and magnitudes to minimize the dispersion between different models by forcing models to match the observed location in the solar-metallicity cluster M67. Even for these calibrated models, however, the internal consistency between models deteriorates at higher metallicities, and photometric masses become smaller than asteroseismic masses, as seen from metal-rich field RC stars with {\it Gaia} parallaxes. Similarly, the average photometric mass for metal-rich NGC~6791 stars ranges from $0.7\ M_\odot$ to $1.1\ M_\odot$, depending on the specific set of models employed. An ensemble average of the photometric masses ($0.88\pm0.16\ M_\odot$) in NGC~6791 is marginally consistent with the asteroseismic mass ($1.16\pm0.04\ M_\odot$). There is a clear tension between the masses that one would predict from photometry for metal-rich field RC stars, asteroseismic masses, and those that would be expected from the ages of stars in the Galactic disk populations and canonical RGB mass loss. We conclude that standard RC models need to be reexamined in light of these powerful new data sets.

\end{abstract}


\section{Introduction}\label{sec:introduction}

The horizontal branch (HB) is a mass sequence in which more massive, helium core-burning stars tend to have lower effective temperatures ($\teff$) or redder colors with a mild increase in luminosity \citep[e.g.,][]{iben:70}. Theoretical models suggest that this monotonic behavior of HB stars extends to $\sim0.7\ M_\odot$. For more massive stars, the luminosity begins to rise steeply because of the increased luminosity from the hydrogen-burning shell, and the $\teff$ trend with mass becomes eventually reversed \citep[e.g.,][]{girardi:99}. At even higher masses ($\ga1.5\ M_\odot$), the luminosity drops precipitously with mass due to smaller core masses until the progenitors of HB stars ($\sim2\ M_\odot$) have large enough masses to ignite helium in a nondegenerate medium. Because the mass gradation along the HB becomes finer toward redder colors, color-magnitude diagrams (CMDs) of old open clusters often show a clustering of red HB stars with a narrow range of luminosity and temperature, called a red clump (RC), as opposed to the continuous HB sequence frequently observed in Galactic globular clusters \citep[see][for recent reviews on the RC]{girardi:16}.

The RC has been used for deriving distances to key stellar systems in the local universe \citep[e.g.,][among others]{paczynski:98,stanek:98,nataf:13}, and a number of studies attempted to derive a zero-point of the calibration given the far-reaching implications on the cosmic distance scale \citep[e.g.,][]{udalski:00,groenewegen:08}. However, the location of the RC depends on the distinct properties of stars in terms of initial mass (or age) and chemical compositions and therefore different star-forming histories of a system \citep[e.g.,][]{girardi:01}. The RC's position on CMDs can also depend on the amount of mass loss on the red giant branch (RGB), which may or may not be correlated with age and/or abundances.

The HB stars have lower masses than their progenitor stars, since stars lose mass while ascending to the tip of the RGB. The amount of the RGB mass loss in clusters can be directly estimated by comparing masses on the HB and main-sequence turn-off (MSTO), from which it has been known for a while that stars in Galactic globular clusters typically lose $\sim25\%$ of their initial mass while ascending the RGB ($\Delta M \sim 0.2\ M_\odot$). The amount of mass loss in globular clusters can be parameterized using the Reimers' formulation \citep{reimers:75} with a dimensionless mass-loss efficiency parameter ($\eta \sim 0.5$), and seems to depend only weakly on metallicity among globular clusters \citep[e.g.,][]{mcdonald:15}.

In this sense, the observed HB morphology in NGC~6791 is puzzling and difficult to explain. Along with RC giants, the cluster harbors about a dozen extreme (or extended) HB (EHB) stars, making its HB morphology unique. These hot ($\teff > 25,000$~K) helium core-burning objects have masses around $\sim0.5\ M_\odot$ with extremely thin hydrogen envelopes \citep{kaluzny:92,liebert:94,kaluzny:95,carraro:13}. The original mass of these stars can be inferred from the mass at the bottom of the RGB; \citet{brogaard:11,brogaard:12} found $1.15\pm0.02\ M_\odot$ based on the mass of the primary in the V20 eclipsing binary system ($1.0868\pm0.0039\ M_\odot$), which is located on the cluster's MSTO. This indicates that progenitors of EHB stars have lost a significant fraction of their initial mass ($\Delta M \ga 0.6\ M_\odot$). In addition to EHB stars, a large population of low-mass white dwarfs (WDs) are found in NGC~6791 \citep{kalirai:07}. The masses of these WDs are $\sim0.43\ M_\odot$, suggesting that their progenitor stars have failed to ignite helium at the tip of the RGB and likely skipped the core helium-burning phase. \citeauthor{kalirai:07} concluded that they are WDs mostly made of helium, unlike normal carbon-oxygen WDs.

The origin of the enhanced mass loss experienced by the progenitors of EHB and helium WDs in NGC~6791 is still under debate. One of the possibilities is enhanced mass loss from their stellar winds \citep[e.g.,][and references therein]{tripicco:93}. Such an effect, even for single stars, is plausible in supersolar-metallicity stars with increased atmospheric opacities, although \citet{vanloon:08} found no evidence of mid-infrared (IR) excess around giant stars in the cluster. It is also possible that, instead, these stars have experienced mass transfer to a companion star in a close binary system \citep[e.g.,][]{liebert:94,han:03}. By contrast, the relatively massive RC giants in NGC~6791 are in stark contrast with the enhanced mass loss of progenitors of EHB stars. As discussed below, theoretical models of RC stars permit either high (up to $\Delta M \sim 0.5\ M_\odot$) or low ($\Delta M \la 0.2\ M_\odot$) mass-loss solutions for the RC; this opens up two potential solutions for the origin of the EHB stars. Either all of the core He-burning stars in the cluster have experienced high mass loss from a similar set of processes or there is a bimodal mass-loss process, far more efficient in the EHB precursors than in other core He-burning stars.

Meanwhile, asteroseismic observations of stellar oscillations, or asteroseismology, have become a fundamental tool for inferring stellar masses and radii of field stars \citep[e.g.,][]{pinsonneault:18}, which can shed light on the amount of the RGB mass loss. Time-series photometry produced by the {\it Kepler} space telescope \citep{borucki:10} has enabled asteroseismic determinations of stellar masses and radii of $\sim7,000$ stars through the so-called solar scaling relations. These relate a large frequency separation ($\Delta \nu$) of solar-like oscillations to a mean stellar density \citep{ulrich:86} and a frequency of maximum power ($\nu_{\rm max}$) to the acoustic cutoff frequency, which depends on the surface gravity ($\log{g}$) and $\teff$ of a star \citep{brown:91,kjeldsen:95}. The mass and radius of a star can be derived from these global oscillation properties, if $\teff$ is known. In this way, \citet{miglio:12} found an unexpectedly small RGB mass loss ($\Delta M = 0.09\pm0.05\ M_\odot$) of RC precursors in NGC~6791 from a comparison with seismic masses of RGB stars ($1.23\pm0.02\ M_\odot$) in that cluster \citep[see also][]{basu:11}.

However, the solar scaling relations need a careful check against independent measurements of stellar mass and radius. Evolved RGB stars have a wide range of effective temperature, mass, metallicity, and internal structures that are far from those of the Sun. For example, \citet{epstein:14} computed asteroseismic masses of nine metal-poor red giants in the halo from a direct application of the scaling relations, and found asteroseismic masses that were systematically too high relative to astrophysical priors ($\Delta M = 0.17\pm0.05\ M_\odot$). On the theoretical side, \citet{white:11} evaluated the accuracy of the mean density relation using a grid of stellar models, and inferred corrections to masses for RGB stars on the order of a few to $\sim10\%$ \citep[see also][]{mosser:13,guggenberger:16,sharma:16}.

In addition to a global zero-point shift in the scaling relations for RGB stars, there is strong evidence for a differential offset in the mean density-- $\Delta \nu$ relation for RC giants. \citet{miglio:12} noted a systematic difference in the mapping of a mean density onto the sound-crossing time between RGB and RC stars due to the significantly different thermal structures of these stars \citep[see also][]{sharma:16}. Furthermore, they found that their asteroseismic radii of RC stars in NGC~6791 were below those computed using the luminosity and $\teff$ of a star, while RGB stars are in agreement. The difference is of the order of $5\%$ in radius, or a $2.7\%$ change in $\Delta \nu$.

The $\nu_{\rm max}$ scaling relation has a weaker physical foundation than that for the frequency spacing. A recent work by \citet{belkacem:11} demonstrated that $\nu_{\rm max}$ depends not only on the acoustic cutoff frequency, but also on the convective Mach number and mixing length parameter. In addition, \citet{viani:17} showed that the $\nu_{\rm max}$ relation has a strong dependence on metallicity, and that the difference from the classical scaling relation can be as large as $2\%$ at [Fe/H]$\sim+0.4$ \citep[see also][]{coelho:15}.

To address these problems, \citet{pinsonneault:18} reassessed the asteroseismic mass and radius estimates for stars in the {\it Kepler} field.  They used an empirical normalization of the $\nu_{\rm max}$ zero-point to ensure agreement between fundamental and asteroseismic masses on the RGB in star clusters. They also included theoretically motivated corrections to the $\Delta \nu$ scaling relations accounting for the different structures of RC and RGB stars and performed rigorous tests of systematic and random errors. \citet{pinsonneault:18} found a negligible difference in mass between the RGB and RC in NGC~6791 ($\Delta M=0.02\pm0.05\ M_\odot$). This implies that a relatively large number of stars in the cluster could have kept most of their initial masses, leaving a clear signature on the mass bimodality among post-RGB stars in the cluster.

We therefore believe that it is a good time to critically evaluate theoretical models of the RC. The goal of this work is to provide independent mass estimates of RC giants from CMDs (hereafter photometric mass estimates) and compare them with asteroseismic masses in the field and in the benchmark cluster. Theoretical models predict a relatively steep mass-luminosity relation of the zero-age HB (ZAHB) at $\ga0.7\ M_\odot$ (see Appendix). Since a cluster's RC can be easily identified on a CMD with its distinct colors and magnitudes, its average mass can be tightly constrained if there exists useful information on distance and reddening along with an accurate metallicity measurement for the cluster. The same is true for field RC giants, in light of {\it Gaia} parallaxes, except that it is far more difficult to distinguish RC giants from the first-ascent giants on a CMD. Fortunately, asteroseismic observations further make it possible to remove a degeneracy and help to select a clean sample of RC giants.

This paper is organized as follows. A field star sample with asteroseismic masses and {\it Gaia} parallaxes is presented in \S~\ref{sec:apokascgaia}. In \S~\ref{sec:sample}, we summarize fundamental cluster parameters of benchmark systems (M67 and NGC~6791), and compare the absolute magnitudes of the cluster RC stars with those of field giants based on {\it Gaia} parallaxes. In \S~\ref{sec:comp} we describe the mapping of photometric data onto mass from a number of theoretical ZAHB models and provide photometric mass estimates for RC stars. In \S~\ref{sec:discussion} we summarize our results. Additional comparisons of theoretical HB models can be found in the Appendix.

\section{APOKASC-{\it Gaia} Sample}\label{sec:apokascgaia}

For our comparison with the photometric mass estimates of RC giants, we utilized asteroseismic masses in \citet{pinsonneault:18}. They published a catalog of fundamental properties of $6676$ giants based on time-series photometry from the {\it Kepler} telescope as a part of the {\it Kepler} Asteroseismic Science Consortium (KASC) and spectroscopic observations in the Apache Point Observatory Galactic Evolution Experiment \citep{apogee}. This joint survey, named APOKASC-2, has classified stars from asteroseismology and produced a list of asteroseismic masses and radii of $\sim2000$ RC giants in the {\it Kepler} field. Typical errors in mass for these stars are about $9\%$ (random) and $8\%$ (systematic), respectively.

We combined the APOKASC-2 catalog with parallax measurements in the second data release (DR2) of the {\it Gaia} mission \citep{gaiadr2}. We used this combined catalog (hereafter APOKASC-{\it Gaia}) to check absolute magnitudes of RC stars from star clusters with independent distance estimates (\S~\ref{sec:sample}). We corrected {\it Gaia} parallaxes in the {\it Kepler} field by adding a parallax zero-point ($0.0528$~mas), as suggested by \citet{zinn:19}. The size of the correction is of the same order of magnitude as an all-sky global zero-point error, $0.029$~mas \citep{lindegren:18}, but it accounts for spatial variations in the zero-point offset specific to the {\it Kepler} field. Our main sample includes $1695$ RC giants with accurate distance measurements ($\sigma_\pi/\pi < 0.1$) within a heliocentric distance $d_{\rm sun} < 5$~kpc. We excluded the so-called secondary RC stars, which have not experienced a helium flash at the tip of the RGB \citep[e.g.][]{girardi:99}, by taking stars with asteroseismic masses $M_{\rm seis} < 2\ M_\odot$.

\citet{pinsonneault:18} adopted theoretical corrections on $\Delta \nu$ on a star-by-star basis that depend on mass, radius, and chemical abundance. Theoretical corrections on the $\nu_{\rm max}$ scaling relation were not applied, as there is no consensus model at the present time. Instead, they derived an effective solar $\nu_{\rm max}$ by requiring that the scaling relations give the same RGB mass as those obtained from eclipsing binary stars near the MSTO in NGC~6791 and NGC~6819. Their corrected scaling relations provide consistent radii for both RGB and RC stars with those from CMDs in these two clusters.

Because the {\it Kepler} field is located in a low Galactic latitude region ($5\deg \la b \la 21\deg$), foreground dust extinction is generally large and spatially variable and is often an important contributor in the error budget for absolute magnitudes of stars. We adopted the \citet{pinsonneault:18} extinction estimates in the APOKASC-2 catalog for individual stars. In brief, they obtained $A_V$ on a star-by-star basis by comparing optical and near-infrared photometry with absolute magnitudes of stars assuming the extinction curve in \citet{cardelli:89} with $R_V\equiv A_V/E(B\, -\, V)=3.1$. They estimated the luminosity of a star from an asteroseismic radius and spectroscopic abundances, and computed absolute magnitudes in each passband by adopting bolometric corrections in the ATLAS9 synthetic spectral library \citep[see also][]{rodrigues:14}. We converted $A_V$ in the APOKASC-2 catalog into $\ebv$ by taking $R_V=3.1$.

Throughout the paper, we adopt extinction coefficients in $BVI_CJHK_s$ as prescribed in \citet{an:07a}, which include values for $R_V$\footnote{We note that color-dependent extinction coefficients were originally derived by \citet{bessell:98} based on the $R_V=3.1$ extinction curve.}, $R_{VI}\equiv E(V\,-\,I_C)/\ebv$, $R_{VK}\equiv E(V\,-\,K_s)/\ebv$, $R_{JK}\equiv E(J\,-\,K_s)/\ebv$, and $R_{HK}\equiv E(H\,-\,K_s)/\ebv$. The prescription explicitly includes color terms, of which the effects are $\sim1\%$ levels in $V$, $K_s$, $\bv$, and $\vi$ at $\ebv\sim0.1$, but negligible in $\jk$ and $\hk$. In the $gri$ passbands, we adopted the constant extinction coefficients with respect to $A_V$ ($A_g/A_V$, $A_r/A_V$, and $A_i/A_V$) in \citet{an:09}, which have been derived using the theoretical stellar spectra of a Sun-like star. There is only limited information available on the color dependence of these values in $ugriz$. We assumed a $10\%$ error in each of the extinction coefficients.

\begin{figure}
\epsscale{1.2}
\plotone{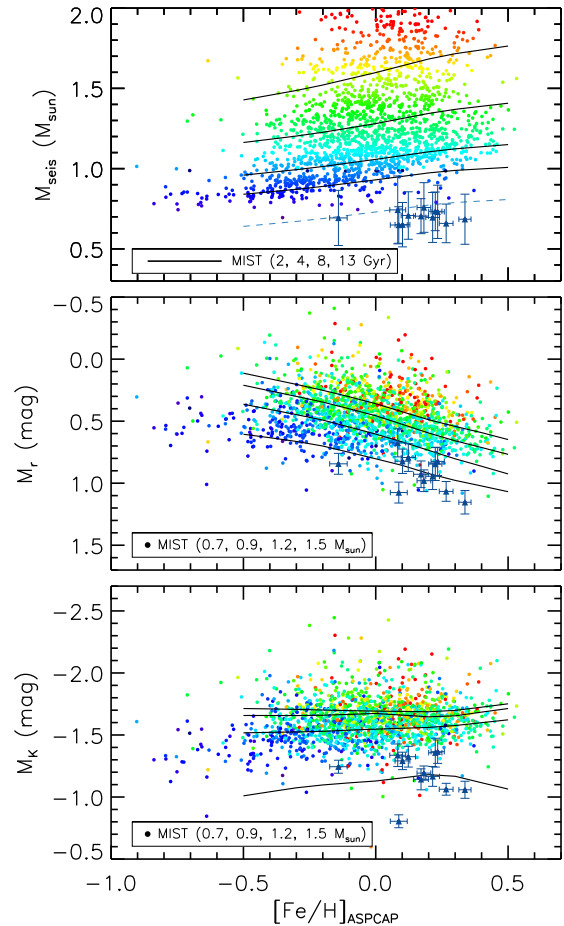}
\caption{The APOKASC-{\it Gaia} sample. Top: asteroseismic masses (shown by different colors) as a function of [Fe/H]. For reference, solid lines represent masses at the tip of the RGB in the MIST models at several different ages. Blue triangles with error bars indicate stars with masses that are at least $0.2\ M_\odot$ smaller than the $13$~Gyr model (dashed line). Middle and bottom: absolute magnitudes in $r$ and $K_s$, respectively, based on {\it Gaia} parallaxes and foreground extinction estimates in the APOKASC catalog. Solid lines are MIST models for a number of ZAHB masses with zero-point color corrections (see text).\label{fig:mag}} \end{figure}

In the top panel of Figure~\ref{fig:mag}, asteroseismic masses of the APOKASC-{\it Gaia} sample are shown by different colors as a function of [Fe/H]. For reference, ZAHB models from the Modules for Experiments in Stellar Astrophysics (MESA) Isochrones \& Stellar Tracks \citep[MIST;][]{choi:16,dotter:16} are overlaid by solid lines at different ages. These models show stellar masses at the tip of the RGB. In other words, their masses indicate upper limits on the RC mass, which become smaller for older ages due to a lower MSTO mass. According to these models, the APOKASC sample seems to cover a wide range of ages from a few Gyr to $13$~Gyr, if one assumes mild mass loss after the helium flash. There are a dozen low-mass stars ($\la 0.8\ M_\odot$; blue triangles with error bars) above solar metallicity. The masses of these stars are even smaller than those predicted by the $13$~Gyr old MIST ZAHB, suggesting enhanced mass loss.\footnote{We note that the MIST isochrones adopt a Reimers constant of $0.1$ based on matching the initial mass-final mass relationship for WDs, which is about three to four times smaller than the value used for globular clusters \citep[e.g.,][]{mcdonald:15}.}

The bottom two panels in Figure~\ref{fig:mag} show absolute magnitudes of the APOKASC-{\it Gaia} stars in $M_r$ and $M_K$, respectively. In these panels, the solid lines represent absolute magnitudes predicted by MIST models at several different ZAHB masses. The observed distribution in $M_r$ indicates that RC giants generally become fainter at higher metallicities, but it is also a strong function of RC mass. More massive stars are brighter than less massive ones, which is consistent with theoretical predictions from the MIST and all other published models. On the other hand, absolute magnitudes in $M_K$ are insensitive to stellar mass and metallicity. 

\begin{figure}
\epsscale{1.2}
\plotone{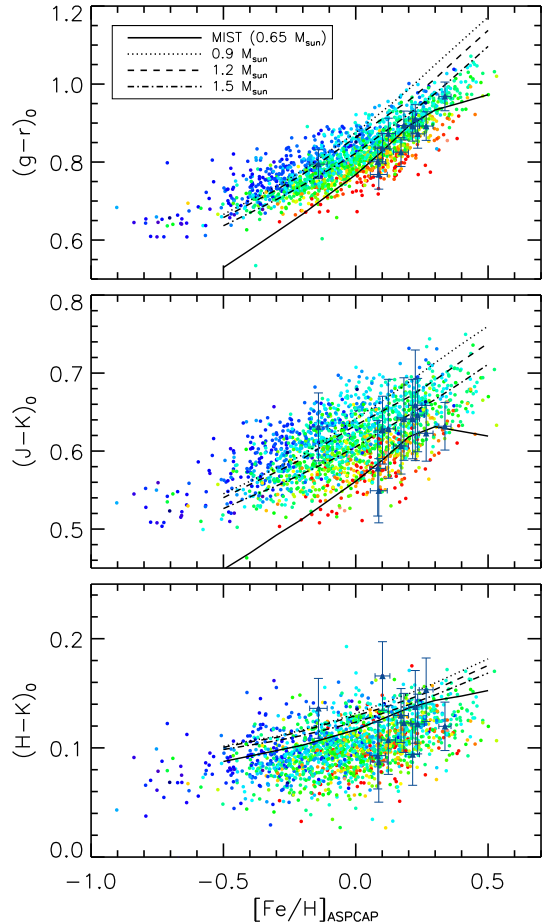}
\caption{Same as in the bottom two panels of Figure~\ref{fig:mag}, but shown in some selected color indices. Note that the ZAHB mass-color relations in the models are reversed below the $0.9\ M_\odot$ model.\label{fig:colors}} \end{figure}

The color distributions of the APOKASC-{\it Gaia} sample are shown in Figure~\ref{fig:colors} for some selected color indices ($\gr$, $\jk$, and $\hk$). Similar to the absolute magnitudes of these stars, the optical colors show a strong dependence on both mass and metallicity, but the infrared colors have significantly less sensitivity. However, the colors of RC giants do not monotonically change with stellar mass, as predicted by modern stellar models. This is exemplified by MIST models with different ZAHB masses. At a given metallicity, less massive RC giants become redder, but the trend is reversed below $\sim 0.9\ M_\odot$. This is also seen in the low-mass RC giants in the APOKASC-{\it Gaia} sample (triangles with error bars). Their masses are less than $0.8\ M_\odot$, but their colors are similar to those of more massive ($\sim1.3\ M_\odot$) RC giants. This implies that colors cannot solely be used to constrain the masses of RC giants.

Previously, \citet{chen:17} estimated the mean absolute colors and magnitudes of RC giants in various photometric passbands based on the {\it Kepler} asteroseismic data. They obtained $\langle M_r \rangle=0.42\pm0.11$ and $\langle M_K \rangle=-1.63\pm0.06$ independent of the {\it Hipparcos} parallax measurements.  Although individual masses and metallicities are not readily available in \citet{chen:17}, their values are broadly consistent with those of our APOKASC-{\it Gaia} sample.

\section{Absolute Magnitudes of RC Giants in Open Clusters}\label{sec:sample}

Our photometric mass estimate of RC giants is based on a direct comparison of observed magnitudes with HB models (\S~\ref{sec:comp}), and therefore requires accurate distance and foreground reddening estimates. While the APOKASC-{\it Gaia} sample provides useful information on the RC luminosity, open clusters are particularly of interest, because the distance and reddening can be determined simultaneously and they are often more precise than those of field stars.

\subsection{Cluster Sample}\label{sec:cluster_sample}

Among old open clusters in the Milky Way, we selected M67 (NGC~2682) and NGC~6791, which show a prominent RC on their CMDs. These two clusters have been sufficiently well studied, and there are extensive data on photometry, metallicity, membership, and useful constraints from eclipsing binaries. Both clusters have been observed by the {\it Kepler}, and the mass of RC stars has been determined through asteroseismic measurements \citep{miglio:12,pinsonneault:18}. In this work, we use the solar-metallicity cluster M67 \citep[${\rm [Fe/H]}=0.00\pm0.01$;][and references therein]{an:07b,onehag:14} as a control sample to derive first-order corrections on HB models. Because NGC~6791 is the most metal-rich \citep[${\rm [Fe/H]}=+0.37\pm0.07$;][and references therein]{an:15} and oldest known cluster \citep[$9\pm1$~Gyr;][and references therein]{brogaard:12,an:15}, RC masses from the photometric and asteroseismic approaches can be compared with each other in the high-metallicity regime \citep[see also][]{pinsonneault:18}.

However, there is a distinct difference between M67 and NGC~6791. The former is $4\pm0.5$~Gyr old \citep{vandenberg:04,bellini:10}, while age estimates for NGC~6791 range from $7$~Gyr to $12$~Gyr \citep[e.g.,][]{chaboyer:99,salaris:04,carney:05,king:05,carraro:06,anthony:07,claret:07,kalirai:07,montalto:07,brogaard:12,an:15}. The average of the two most recent age estimates from MSTO \citep{brogaard:12,an:15} is $9\pm1$~Gyr, which is adopted in the following analysis. The progenitor mass of RC giants is set by a stellar age and therefore is significantly smaller in NGC~6791 than in M67. An important consequence of this is a systematically fainter luminosity of RC giants in NGC~6791. Below, we derive the absolute magnitudes and extinction-corrected colors of RC giants in M67 and NGC~6791, and compare them with those of field stars in the APOKASC-{\it Gaia} sample.

\subsection{Distance and Foreground Extinction}\label{sec:param}

In the following analysis, we adopted the best available set of cluster distance and reddening in the literature. There are two eclipsing binary studies in M67 \citep{yakut:09,gokay:13}, which found $\dmn=9.66\pm0.09$ and $\dmn=9.56\pm0.06$, respectively. Meanwhile, \citet{an:07b} derived $\dmn=9.61\pm0.03$ from MS fitting using a set of isochrones with empirically calibrated colors. All of these estimates are in good agreement with each other, and their weighted mean distance modulus yields $\dmn=9.61\pm0.03$. As an alternative check on these results, we took likely cluster members from CMDs in \citet{sandquist:04} and searched for their {\it Gaia} parallaxes. By directly inverting parallaxes, we computed a mean distance modulus of the cluster $\dmn=9.66\pm0.13$ after correcting for the global parallax zero-point offset ($0.029$~mas) as suggested by the {\it Gaia} team \citep{lindegren:18}. The above error in distance includes errors propagated from individual parallax errors or a standard deviation, whichever is larger, and an error of $0.04$~mas, added in quadrature, to represent the effect of spatial correlations in {\it Gaia} parallaxes over angular scales of about $1\arcdeg$ \citep{lindegren:18}. The {\it Gaia} distance is in good agreement with other distance estimates, but the mean distance modulus has a relatively large error. Therefore, we adopted the above weighted mean distance (without {\it Gaia}'s) in the following analysis. For the foreground reddening, we adopted $\ebv = 0.041\pm0.004$, which is a weighted mean from previous estimates in the literature and a solution based on the calibrated isochrones \citep{an:07b}.

For NGC~6791, \citet{brogaard:11,brogaard:12} determined accurate astrophysical parameters for two eclipsing binary systems (V18 and V20) in the cluster. Primary and secondary stars in these binary systems are on the MS, and their masses and radii were determined with errors of less than $1\%$. By combining stellar radii with spectroscopic temperatures, they found $\dmnv=13.51\pm0.06$ for a mean apparent distance modulus of NGC~6791. \citeauthor{brogaard:12} also employed theoretical stellar models to find $\ebv=0.14\pm0.02$ from a $VI$ CMD of the cluster, which results in $\dmn=13.08\pm0.09$. Meanwhile, \citet{an:15} provided $\dmn=13.04\pm0.08$ and $\ebv=0.105\pm0.014$ based on MS fitting of the calibrated isochrones. We took the average values from these two studies: $\dmn = 13.06\pm0.06$ and $\ebv = 0.12\pm0.02$. In addition, we computed a mean distance modulus $\dmn=13.3\pm0.8$ from {\it Gaia} parallaxes for asteroseismic members on the cluster's RGB in \citet{stello:11}. The constraint from the {\it Gaia} is weak, because its parallax is overwhelmed by a spatially correlated error \citep{lindegren:18}.

\subsection{Cluster Photometry}\label{sec:photometry}

We utilized cluster photometry in various photometric passbands: $gri$ in the Sloan Digital Sky Survey \citep[SDSS;][]{dr7}, $BVI_C$ in Johnson-Cousins, and $JHK_s$ in the Two Micron All Sky Survey \citep[2MASS;][]{skrutskie:06}. These passbands cover a wide range of wavelength and show different degrees of sensitivity on metallicity and reddening, providing a useful check on underlying systematic errors in our RC mass estimates. We excluded SDSS $u$ and Johnson $UR_C$ passbands because of large systematic errors in the photometry and models \citep[e.g.,][]{an:08}. We also did not use the $z$-band data, because their photometry is not available in M67, and therefore the M67-based model calibration could not be performed (see below).

We utilized $gri$ photometry of NGC~6791 from \citet{an:08} that has been extracted from the original SDSS imaging data using the DAOPHOT/ALLFRAME suite of programs \citep{stetson:87,stetson:94}. We used the average magnitudes and colors from three SDSS imaging runs ($5403$, $5416$, $6177$) and adjusted the photometric zero-points from \citet{an:08} to put them on the {\tt UberCal} system \citep{padmanabhan:08}. We note that the corrections amount to only a few thousandth magnitudes \citep{an:13}. While RC stars in NGC~6791 have been directly observed in SDSS, RC giants in M67 were too bright and saturated in the SDSS images because of the maximum brightness limit in the survey ($r\sim14$~mag). Instead, we utilized $g'r'i'$ photometry for secondary cluster standard stars from J.\ Clem \citep[see also][]{clem:07}\footnote{See http://www.phys.lsu.edu/~jclem/research/photometry.html}, and transformed it into the natural SDSS system using the transformation equations in \citet{tucker:06}. We took the average $BVI_C$ data for RC giants in M67 from \citet{montgomery:93} and \citet{sandquist:04}. For NGC~6791, we took the updated version of the \citet{stetson:03} photometry.\footnote{See http://www3.cadc-ccda.hia-iha.nrc-cnrc.gc.ca/\\community/STETSON/homogeneous/NGC6791/. The updated photometry was first referenced in \citet{brogaard:12}.} We used $JHK_s$ photometry from the All Sky Data Release of the 2MASS Point Source Catalog (PSC)\footnote{See http://www.ipac.caltech.edu/2mass/.} after removing objects that are undetected, confused, and/or blended.

\begin{figure*}
\epsscale{0.95}
\plotone{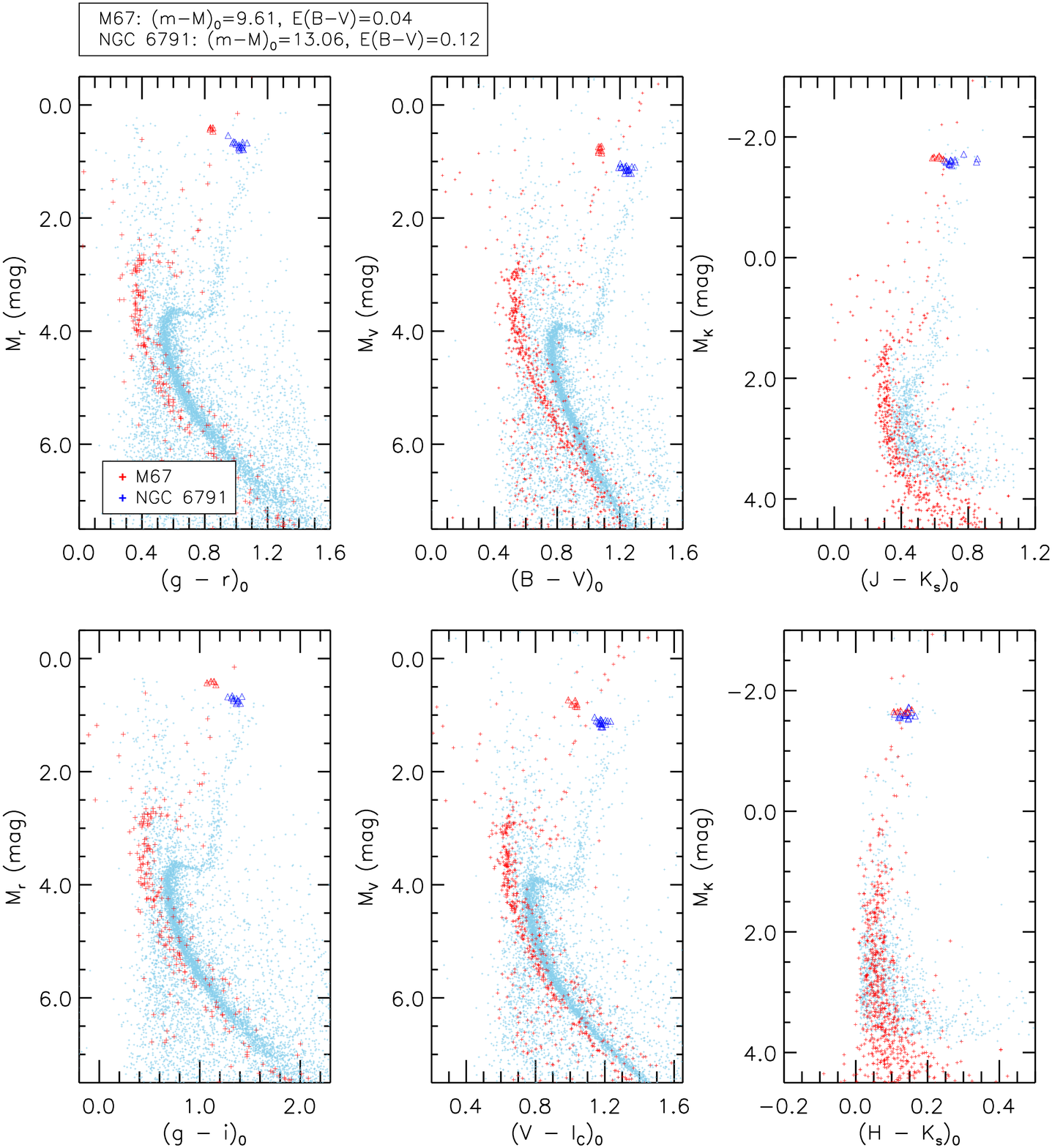}
\caption{The $griBVI_CJHK_s$ CMDs of M67 \citep[][red points]{montgomery:93,clem:07} and NGC~6791 \citep[][blue points]{stetson:03,an:15} with absolute magnitudes and reddening-corrected colors. Asteroseismic members of the cluster's RC are highlighted by open triangles \citep{stello:11,stello:16}. For NGC~6791, transformed $JHK_s$ data from \citet{carney:05} are shown in order to depict cluster sequences beyond the completeness limit of 2MASS.\label{fig:cmd}} \end{figure*}

Figure~\ref{fig:cmd} displays the CMDs of M67 and NGC~6791 with absolute magnitudes and reddening-corrected colors in $griBVI_CJHK_s$. The photometry of M67 and NGC~6791 is shown by red and blue points, respectively. Because the NGC~6791 field is crowded with a significant amount of foreground/background contamination at low Galactic latitudes, only stars within a $10\arcmin$ radius from the cluster's center are displayed. For NGC~6791, the $JHK_s$ data in \citet{carney:05} are shown in Figure~\ref{fig:cmd} in order to show the cluster's faint sequence beyond the completeness limit in 2MASS.

\subsection{RC Sample Selection}\label{sec:selection}

The localized RC feature on CMDs is clearly distinguishable from the RGB in both clusters: $0.8 < \gr < 0.9$ and $10.0 < r < 10.3$ in M67 and $1.05 < \gr < 1.20$ and $14.0 < r < 14.2$ in NGC~6791. However, we took the RC giants identified from the previous asteroseismic studies \citep{stello:11,stello:16} for the following analysis. In total, there are seven and $19$ asteroseismic RC members in M67 and NGC~6791, respectively, which are highlighted by red and blue triangles in Figure~\ref{fig:cmd}. \citet{sandquist:04} identified four of the stars as RC members of M67 based on both photometry and proper-motion membership information in \citet{sanders:77} and \citet{girard:89}. In addition, the heliocentric radial velocities of the seven stars in M67 have $+34 \la v_r \la +35$~km\ s$^{-1}$ \citep{pasquini:11,reddy:13}, which are reasonably close to the mean heliocentric radial velocity of the cluster \citep[$\langle v_r \rangle = +33$~km\ s$^{-1}$;][]{lee:08}. Most of the RC giants in \citet{stello:11} have been identified as members of NGC~6791 based on proper-motion and radial velocity measurements in the WIYN Open Cluster Study \citep{platais:11,tofflemire:14}.

\begin{deluxetable}{lcc}
\tablewidth{0pt}
\tablecaption{Observed Mean Magnitudes and Colors of RC Giants in M67 and NGC~6791 \label{tab:obsmag}}
\tablehead{
  \colhead{Magnitudes/} &
  \multicolumn{2}{c}{Cluster} \\
  \cline{2-3}
  \colhead{Colors} &
  \colhead{M67} &
  \colhead{NGC~6791}
}
\startdata
$r$   & $10.151\pm0.015$ & $14.110\pm0.009$ \\
$\gr$ & $ 0.887\pm0.005$ & $ 1.137\pm0.006$ \\
$\gi$ & $ 1.192\pm0.019$ & $ 1.560\pm0.008$ \\
\cline{1-3}         
$V$   & $10.519\pm0.012$ & $14.571\pm0.011$ \\
$\bv$ & $ 1.107\pm0.005$ & $ 1.349\pm0.006$ \\
$\vi$ & $ 1.075\pm0.005$ & $ 1.329\pm0.005$ \\
\cline{1-3}
$K_s$ & $ 7.959\pm0.008$ & $11.504\pm0.012$ \\
$\jk$ & $ 0.643\pm0.011$ & $ 0.763\pm0.007$ \\
$\hk$ & $ 0.135\pm0.013$ & $ 0.161\pm0.007$
\enddata
\end{deluxetable}

The mean RC magnitudes and colors are presented in Table~\ref{tab:obsmag}. Errors represent a standard error of the mean (s.e.m.) from the scatter of RC giants on CMDs. If there is more than a single source of photometry ($BVI$ for M67 and $gri$ for NGC~6791), we took a weighted standard deviation of an RC's mean position or a propagated error from the scatter, whichever is larger. The size of the error listed in Table~\ref{tab:obsmag} is typically $0.01$~mag, but additional systematic errors in photometry could be of the same order as follows. For the $gri$ photometry of M67, we checked the photometric transformation errors by comparing the colors and magnitudes of unsaturated stars (fainter than the cluster's RC and $r < 20$~mag) in \citet{clem:07} with those in \citet{an:08}, and found differences of $0.010$~mag, $0.022$~mag, $0.028$~mag in $r$, $\gr$, and $\gi$, respectively. For the $gri$ photometry in NGC~6791, we found that the differences between (transformed) \citet{clem:08} and \citet{an:08} photometry (run 5416) are $0.017$~mag, $0.006$~mag, $0.009$~mag in $r$, $\gr$, and $\gi$, respectively. Given this, we adopted a uniform $2\%$ error in photometry in all passbands in the following analysis.

Unlike M67, NGC~6791 exhibits differential foreground dust extinction across the field \citep[e.g.,][]{platais:11,brogaard:12}. We used an extinction map in \citet{brogaard:12} to measure the amount of differential reddening for our RC sample.  Among $19$ RC stars included in this study, such measurements are available for $16$ stars. However, we found that an average of the differential reddening is negligible for these stars, $\langle \Delta \ebv \rangle=0.001$ with $\sigma=0.008$.

It may be that the cluster's RC represents an evolutionary channel formed by a small group of stars in the cluster and that their distinct properties, such as the reduced (or enhanced) helium abundance, would make them systematically fainter (or brighter), leading to apparently lower (or larger) photometric masses. Indeed, the notion that NGC~6791 may harbor multiple stellar populations like other Galactic globular clusters \citep[see][and references therein]{gratton:12} has been obtained from the inspection of its CMD. Using stars near the MSTO region, \citet{twarog:11} argued for an extended star-forming history of the cluster over $\sim1$~Gyr. The width of the cluster's RGB has also been known to be too large compared to the size of the photometric errors, even in a CMD with stars selected from proper-motion measurements \citep{platais:11}. However, the variable reddening across the field \citep{brogaard:12} is still one of the major uncertainties, which prevents identifying multiple stellar sequences on CMDs.

\citet{geisler:12} claimed to have discovered two chemically distinct groups of stars in NGC~6791 in the [Na/Fe]-versus-[O/Fe] plane, which would have been similar to the chemical inhomogeneities of light elements found in globular clusters \citep[e.g.,][]{gratton:14}. We identified six stars in \citet{geisler:12} as RC giants based on their positions on the CMDs. However, they are equally split into the above two chemically distinct groups. The existence of such mixed populations is contradictory to a narrow range in temperature and luminosity of RC stars. Furthermore, diverse Na abundances could not be confirmed in more recent, high-resolution $H$-band spectroscopic studies \citep{bragaglia:14,cunha:15}.

\subsection{Absolute Magnitudes, Colors, and Their Errors}

\begin{figure*}
\epsscale{0.95}
\plotone{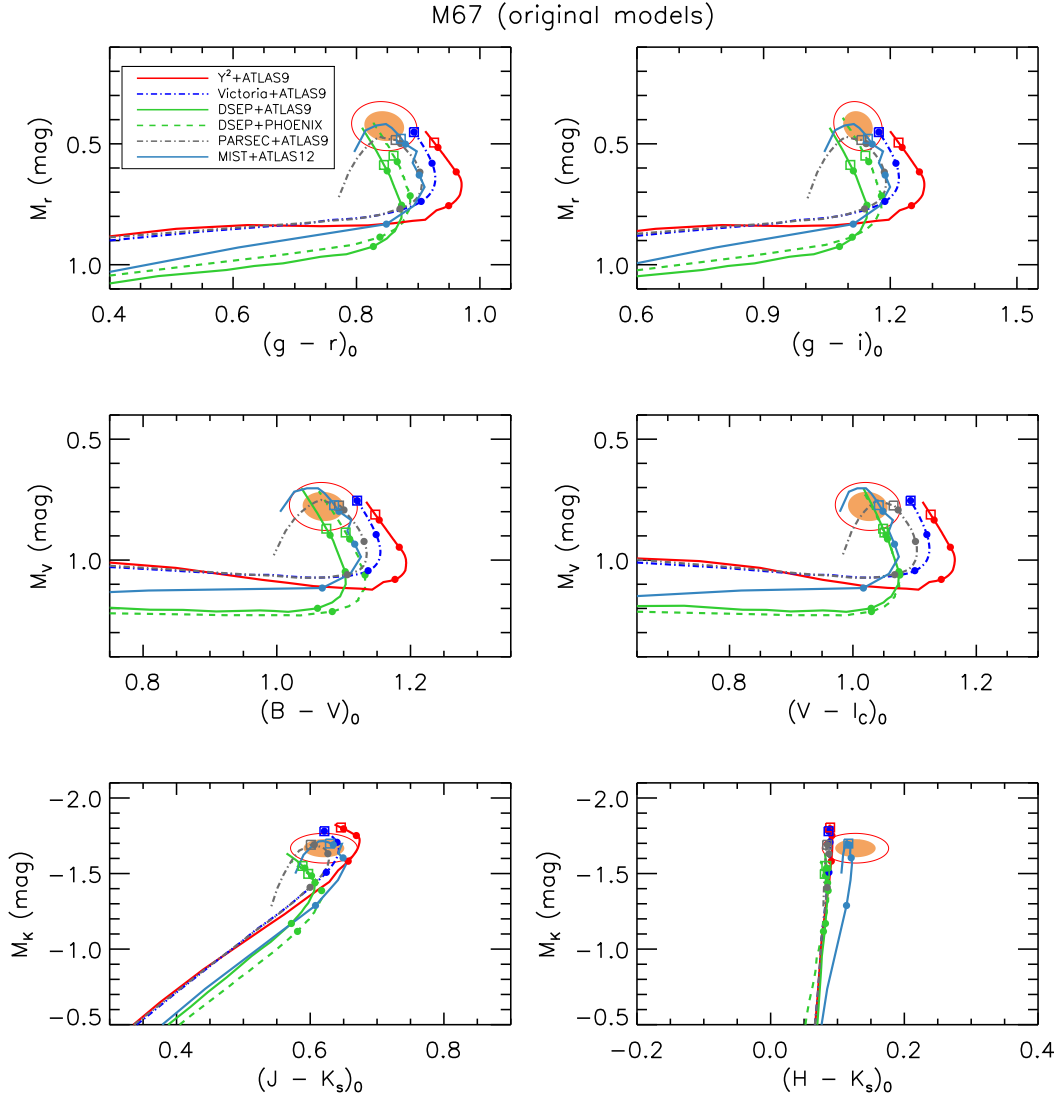}
\caption{Mean absolute magnitudes and reddening-corrected colors of RC giants in M67 in the SDSS, Johnson-Cousins, and 2MASS photometric systems. Their $68\%$ and $95\%$ confidence intervals are shown by shaded and solid ellipses, respectively. Theoretical ZAHB models at solar metallicity are displayed in various combinations of interior models and color transformations (see \S~\ref{sec:models}). Along each grid line, $0.7\ M_\odot$, $1.0\ M_\odot$, and $1.3\ M_\odot$ models are marked by filled circles. The position of a $1.36\ M_\odot$ model is additionally shown by an open square, except that the point enclosed by a square for the Victoria+ATLAS9 model corresponds to its upper mass limit ($1.3\ M_\odot$).\label{fig:comp67}} \end{figure*}

\begin{figure*}
\epsscale{0.95}
\plotone{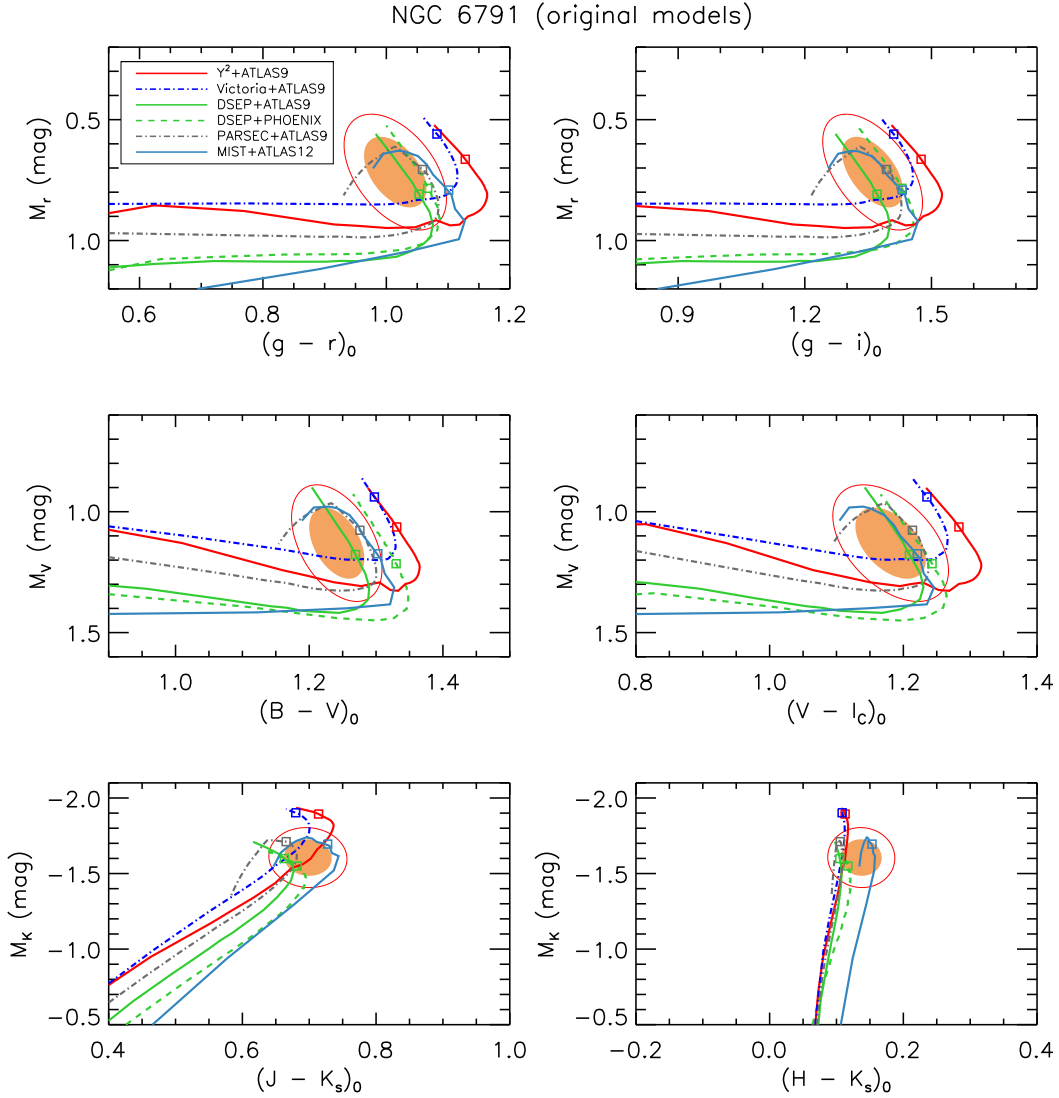}
\caption{Same as Figure~\ref{fig:comp67} but displaying confidence intervals of RC giants in NGC~6791. Theoretical ZAHB models are shown at [Fe/H]$=+0.37$ without zero-point corrections (see \S~\ref{sec:m67}). Along each model grid, the position of a $1.15\ M_\odot$ model is marked by an open square.\label{fig:comp6791a}} \end{figure*}

\begin{deluxetable}{lcc}
\tablewidth{0pt}
\tablecaption{Absolute Magnitudes and Extinction-corrected Colors of RC Giants in M67 and NGC~6791 \label{tab:absmag}}
\tablehead{
  \colhead{Magnitudes/} &
  \multicolumn{2}{c}{Cluster} \\
  \cline{2-3}
  \colhead{Colors} &
  \colhead{M67} &
  \colhead{NGC~6791}
}
\startdata
$M(r)$                    &  $0.43\pm0.04$ & $0.72\pm0.10$ \\
$(g\, -\, r)_0$           &  $0.85\pm0.02$ & $1.01\pm0.03$ \\
$(g\, -\, i)_0$           &  $1.12\pm0.02$ & $1.36\pm0.05$ \\
\cline{1-3}                  
$M(V)$                    &  $0.78\pm0.04$ & $1.13\pm0.10$ \\
$(B\, -\, V)_0$           &  $1.07\pm0.02$ & $1.24\pm0.03$ \\
$(V\, -\, I_C)_0$         &  $1.02\pm0.02$ & $1.18\pm0.04$ \\
$M(I_C)$\tablenotemark{a} & $-0.25\pm0.05$ & $-0.05\pm0.10$ \\
\cline{1-3}
$M(K_s)$                  & $-1.67\pm0.04$ & $-1.61\pm0.08$ \\
$(J\, -\, K_s)_0$         &  $0.62\pm0.02$ & $0.70\pm0.02$ \\
$(H\, -\, K_s)_0$         &  $0.13\pm0.02$ & $0.14\pm0.02$
\enddata
\tablenotetext{a}{Computed from $M_V$ and $(V\, -\, I_C)_0$.}
\end{deluxetable}

The shaded area in Figure~\ref{fig:comp67} shows a $68\%$ confidence interval, and the solid ellipse indicates a $95\%$ confidence interval of the RC's position in each of the CMDs of M67. Similarly, both the $68\%$ and $95\%$ confidence intervals in NGC~6791 are shown in Figure~\ref{fig:comp6791a}. We estimated the confidence levels by generating absolute magnitudes and extinction-corrected colors of $10^5$ simulated RC giants for which we assumed gaussian errors in the input parameters, such as photometric zero-point offsets, extinction coefficients, and the cluster's metallicity, distance, and reddening. Table~\ref{tab:absmag} lists the absolute magnitudes and colors of RC stars in M67 and NGC~6791 after foreground extinction corrections. Errors represent the $68\%$ confidence intervals of the mean RC positions on CMDs. The $M_I$ magnitudes were computed from $M_V$ and $(\vi)_0$.

\begin{deluxetable*}{lccccc}
\tablewidth{0pt}
\tablecaption{Errors in Absolute Magnitudes and Reddening-corrected Colors of RC Giants\label{tab:rcerr}}
\tablehead{
  \colhead{Magnitudes/} &
  \multicolumn{4}{c}{Source of Error} &
  \colhead{Total} \\
  \cline{2-5}
  \colhead{Colors} &
  \colhead{Scatter} &
  \colhead{Distance} &
  \colhead{Reddening} &
  \colhead{Zero-point\tablenotemark{a}} &
  \colhead{Error}
}
\startdata
\multicolumn{6}{c}{M67} \\
\cline{1-6}
$M(r)$                    & $\pm0.015$ & $\pm0.03$ &  $\pm0.02$ &  $\pm0.02$ &  $\pm0.04$ \\
$(g\, -\, r)_0$           & $\pm0.005$ & \nodata   &  $\pm0.01$ &  $\pm0.02$ &  $\pm0.02$ \\
$(g\, -\, i)_0$           & $\pm0.019$ & \nodata   &  $\pm0.01$ &  $\pm0.02$ &  $\pm0.03$ \\
$M(V)$                    & $\pm0.012$ & $\pm0.03$ &  $\pm0.02$ &  $\pm0.02$ &  $\pm0.04$ \\
$(B\, -\, V)_0$           & $\pm0.005$ & \nodata   &  $\pm0.00$ &  $\pm0.02$ &  $\pm0.02$ \\
$(V\, -\, I_C)_0$         & $\pm0.005$ & \nodata   &  $\pm0.01$ &  $\pm0.02$ &  $\pm0.02$ \\
$M(K_s)$                  & $\pm0.008$ & $\pm0.03$ &  $\pm0.02$ &  $\pm0.02$ &  $\pm0.04$ \\
$(J\, -\, K_s)_0$         & $\pm0.011$ & \nodata   &  $\pm0.00$ &  $\pm0.02$ &  $\pm0.02$ \\
$(H\, -\, K_s)_0$         & $\pm0.013$ & \nodata   &  $\pm0.00$ &  $\pm0.02$ &  $\pm0.02$ \\
\cline{1-6}
\multicolumn{6}{c}{NGC~6791} \\
\cline{1-6}
$M(r)$                    & $\pm0.009$ & $\pm0.06$ &  $\pm0.07$ &  $\pm0.02$ &  $\pm0.10$ \\
$(g\, -\, r)_0$           & $\pm0.006$ & \nodata   &  $\pm0.02$ &  $\pm0.02$ &  $\pm0.03$ \\
$(g\, -\, i)_0$           & $\pm0.008$ & \nodata   &  $\pm0.03$ &  $\pm0.02$ &  $\pm0.05$ \\
$M(V)$                    & $\pm0.011$ & $\pm0.06$ &  $\pm0.07$ &  $\pm0.02$ &  $\pm0.10$ \\
$(B\, -\, V)_0$           & $\pm0.006$ & \nodata   &  $\pm0.03$ &  $\pm0.02$ &  $\pm0.03$ \\
$(V\, -\, I_C)_0$         & $\pm0.005$ & \nodata   &  $\pm0.04$ &  $\pm0.02$ &  $\pm0.04$ \\
$M(K_s)$                  & $\pm0.012$ & $\pm0.06$ &  $\pm0.05$ &  $\pm0.02$ &  $\pm0.08$ \\
$(J\, -\, K_s)_0$         & $\pm0.007$ & \nodata   &  $\pm0.01$ &  $\pm0.02$ &  $\pm0.02$ \\
$(H\, -\, K_s)_0$         & $\pm0.007$ & \nodata   &  $\pm0.01$ &  $\pm0.02$ &  $\pm0.02$
\enddata
\tablenotetext{a}{Uniform photometric zero-point errors.}
\end{deluxetable*}

In Table~\ref{tab:rcerr}, we present contributions of individual errors to the absolute magnitudes and extinction-corrected colors of RC giants in both clusters. The sources of the errors include an observed scatter on CMDs, errors in the cluster's distance and reddening, and photometric calibration and/or transformation errors (denoted by `Zero-point'). Errors in the extinction coefficients are incorporated into the error in reddening. Among these, errors in both distance and reddening dominate the total error budget for absolute magnitudes in NGC~6791. The ongoing {\it Gaia} mission \citep{gaia} will eventually help pin down its distance, but the foreground reddening will remain as a major bottleneck in the determination of its RC luminosity.

\subsection{Comparisons with the APOKASC-{\it Gaia} Sample}

\begin{figure}
\epsscale{1.2}
\plotone{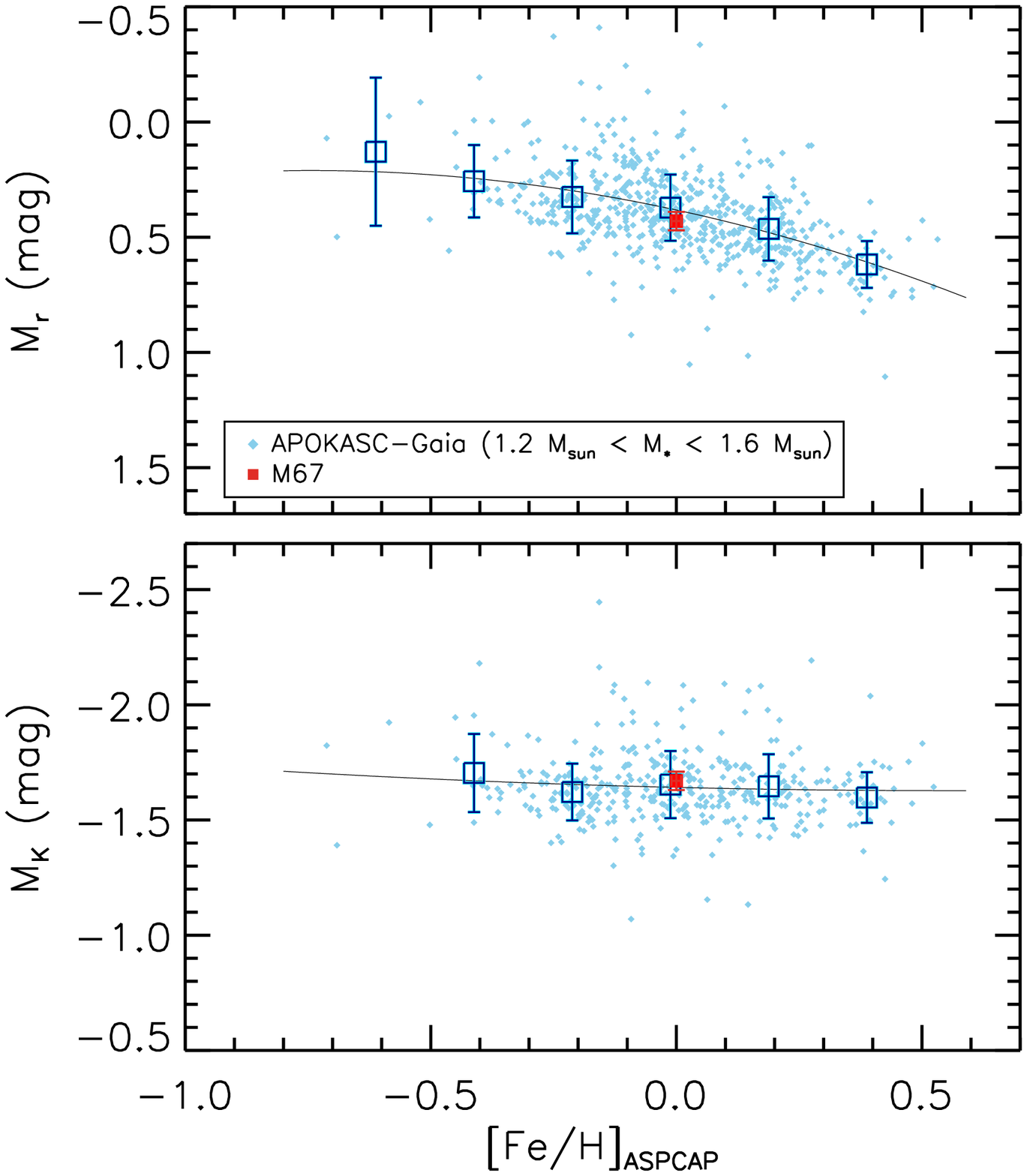}
\caption{Comparison of absolute magnitudes in $r$ (top) and $K_s$ (bottom) between the mean M67 locus and the APOKASC-{\it Gaia} sample ($1.2\ M_\odot < M_* < 1.4\ M_\odot$). For the latter sample, moving box averages and standard deviations in bins of $\Delta$[Fe/H]$=0.2$ are shown by open squares with error bars.\label{fig:m67}} \end{figure}

\begin{figure}
\epsscale{1.2}
\plotone{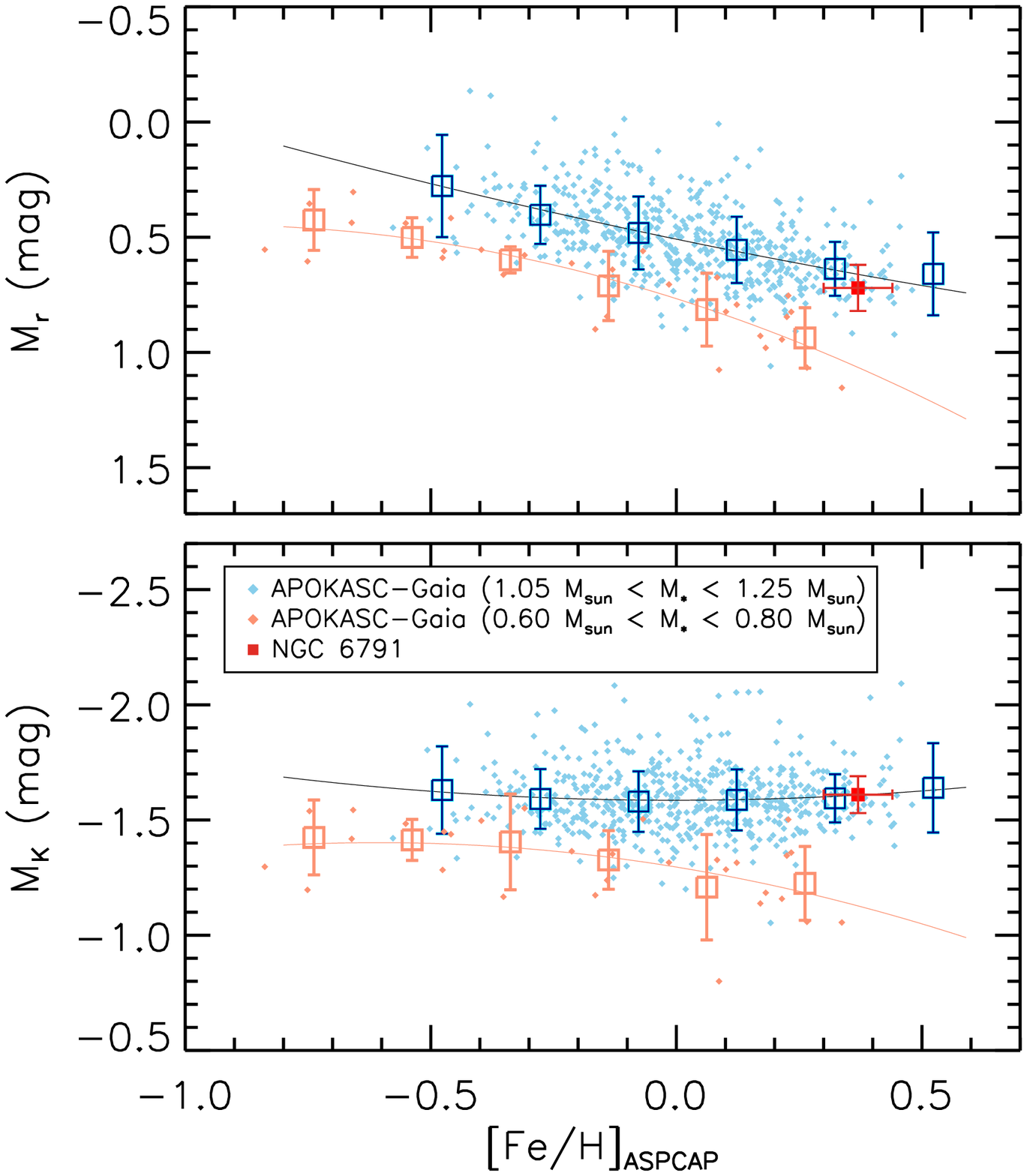}
\caption{Same as Figure~\ref{fig:m67} but comparing absolute magnitudes of lower-mass APOKASC-{\it Gaia} stars with average values of RC stars in NGC~6791.\label{fig:n6791}} \end{figure}

We checked the absolute magnitudes of RC giants in the above two clusters by comparing with those of the APOKASC-{\it Gaia} sample. Figures~\ref{fig:m67}--\ref{fig:n6791} show comparisons in $M_r$ and $M_K$ for M67 and NGC~6791, respectively. These comparisons are independent of theoretical HB models. However, as shown in Figure~\ref{fig:mag}, an absolute magnitude of an RC giant is a strong function of its current mass; therefore, a direct comparison with field RC giants requires an assumption on the range of stellar mass. The progenitor mass of RC giants in M67 is relatively well constrained from the astrophysical prior. Assuming $4.0\pm0.5$~Gyr, the solar-metallicity isochrone from the PAdova and TRieste Stellar Evolution Code \citep[PARSEC;][]{bressan:12} predicts $1.34\pm0.06\ M_\odot$ for an initial mass of the progenitor stars. If the same mass-loss efficiency ($\eta\sim 0.3$) of the \citet{reimers:75} mass-loss formula is taken as in the Galactic globular clusters, the expected amount of mass loss along the RGB is small ($\Delta M \sim 0.05\ M_\odot$)\footnote{Computed using an isochrone generation tool for PARSEC at http://stev.oapd.inaf.it/cgi-bin/cmd, which is maintained at the Osservatorio Astronomico di PARSEC.}. With a similar amount of RGB mass loss, various models employed in this work (see below) agree upon a mass at the tip of the RGB ($1.3$--$1.4\ M_\odot$). The APOKASC-2 catalog does not include stars in M67, but solar-like oscillations of RGB and RC giants in M67 have also been detected from the {\it Kepler} ecliptic mission \citep{stello:16}. An average RGB mass from these seismic measurements is $1.36\pm0.03\ M_\odot$, while an average RC mass is $1.40\pm0.05\ M_\odot$, suggesting little mass loss along the RGB.

In Figure~\ref{fig:m67}, we selected field giants from the APOKASC-{\it Gaia} sample with $1.2\ M_{\rm sun} < M_* < 1.6\ M_{\rm sun}$, covering masses similar to those in M67 from asteroseismology. A second-order polynomial was used to trace the observed magnitude distribution of the field giants, and open squares with error bars indicate average magnitudes and standard deviations in bins of [Fe/H]. At the solar metallicity, our estimated absolute magnitudes of RC giants in M67 are in good agreement with the results from the APOKASC-{\it Gaia} sample.

In Figure~\ref{fig:n6791}, we provided two field giant samples with different mass ranges. Orange points show RC giants with $0.6\ M_\odot < M_* < 0.8\ M_\odot$, indicating low-mass solutions from some of the ZAHB models, as described in the next section. Blue points indicate RC stars with $1.05\ M_\odot < M_* < 1.25\ M_\odot$ based on asteroseismic masses, consistent with the NGC~6791 results for \citet{pinsonneault:18}, who reported a mean RC mass of $1.16\pm0.04\ M_\odot$. Some of the difference in the RC mass between M67 and NGC~6791 reflects a difference in the ages of these clusters, which sets a progenitor's mass for a clump giant. According to PARSEC isochrones ([Fe/H]$=0.37$ and $Y=0.306$), stars with an initial mass of $1.12\pm0.04\ M_\odot$ on the zero-age MS reach the core helium-burning phase at the age of $9\pm1$~Gyr. This estimate is close to the mass at the bottom of the RGB ($1.15\pm0.02\ M_\odot$)\footnote{According to the PARSEC isochrones, the mass of an RC precursor is only $0.01\ M_\odot$ larger than the original mass of stars at the bottom of the RGB at the age of $9$~Gyr.} from the analysis of the eclipsing binary system \citep{brogaard:11,brogaard:12}.

As shown in Figure~\ref{fig:n6791}, our estimated absolute magnitudes of RC stars in NGC~6791 are in better agreement with a subset of the APOKASC-{\it Gaia} stars with $1.05\ M_\odot < M_* < 1.25\ M_\odot$. This suggests that our cluster distance scale is consistent with those based on the {\it Gaia} parallaxes of field stars, and that their masses are also consistent with asteroseismic values. On the other hand, to make the low-mass solution (more severe mass loss) viable, there should be a large systematic error in the asteroseismic mass scale. Because of the good agreement with absolute magnitudes in M67 (Figure~\ref{fig:m67}), this implies that the asteroseismic mass scale is stretched below $\sim1.3\ M_\odot$, while having little or no systematic error at higher masses. We look into this in more detail in the next section by estimating photometric RC masses for the APOKASC-{\it Gaia} sample based on a number of HB models.

\subsection{Comparisons with Previous Studies}

In addition to the APOKASC-{\it Gaia} sample, a number of studies in the literature have reported absolute magnitudes and colors of RC giants based on distances to star clusters or trigonometric parallaxes published before {\it Gaia} DR2. Based on the {\it Hipparcos} parallax measurements, \citet{stanek:98} determined a mean absolute magnitude of field RC stars in the solar neighborhood and found $\langle M (I_C) \rangle=-0.23$ with a dispersion of $\sim0.2$~mag. \citet{paczynski:98} determined $\langle \vi \rangle=1.01$ for the mean color of RC stars in the solar neighborhood. More recently, \citet{hawkins:17} and \citet{ruiz:17} used parallaxes from a joint Tycho-{\it Gaia} astrometric solution \citep{lindegren:16} to obtain $\langle M (K_s) \rangle=-1.61\pm0.01$ and $\langle M (K_s) \rangle=-1.606\pm0.009$, respectively. All of these estimates are consistent with absolute magnitudes of RC stars in M67 (Table~\ref{tab:absmag}). The mean metallicity of stars in the {\it Hipparcos} sample \citep[][$\langle {\rm [Fe/H]} \rangle = -0.12\pm0.01$]{girardi:01} is near solar, so the effect of a metallicity difference from M67 does not likely exceed $\Delta M_I = 0.05$~mag (and less in longer-wavelength bands).

\begin{figure}
\epsscale{1.2}
\plotone{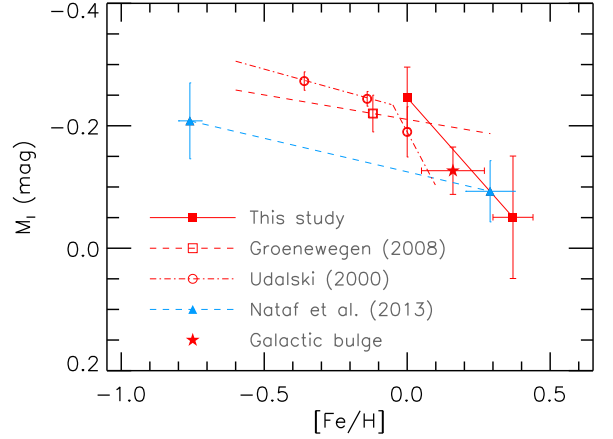}
\caption{Absolute $I_C$-band magnitudes of RC giants in the Milky Way. Red filled squares indicate the mean $M_I$ magnitudes of RC giants in M67 and NGC~6791 from this work, which are connected by a straight line for simplicity. For comparison, absolute magnitudes of nearby field RC giants in the {\it Hipparcos} catalog are shown by red open circles \citep{udalski:00} and an open square \citep{groenewegen:08}. The blue dashed line represents the metallicity sensitivity of the RC luminosity in \citet{nataf:13} based on data from 47~Tuc and NGC~6791 (blue filled triangles). The red star indicates the mean position of RC giants in the Galactic bulge at an average dynamical distance to the Galactic center.\label{fig:rclum}} \end{figure}

Figure~\ref{fig:rclum} displays previous $M_I$ estimates in the literature that contain additional information on metallicity. The red open circles, connected by a red dotted-dashed line, represent mean $I_C$ magnitudes in \citet{udalski:00} in three different metallicity bins. In addition, the mean $I_C$ magnitude of the {\it Hipparcos} RC giants in \citet{groenewegen:08} is shown by a red open square, and its metallicity sensitivity is represented by a red dashed line over an approximate range of metallicity covered by the sample. The red star indicates the mean position of the RC in the Galactic bulge, adopting the dereddened mean $I_C$-band magnitude in \citet[][$I_C=14.443$~mag]{nataf:13} and the distance to the Galactic center, $8.10\pm0.24$~kpc, from the two most recent studies of stellar orbits near Sgr~A$^*$ \citep{chatzopoulos:15,boehle:16}. The Galactic bulge shows a wide range of metallicity, from [Fe/H]$\sim-1.5$ to [Fe/H]$\sim+0.5$, as revealed from comprehensive spectroscopic surveys on microlensed MS dwarfs and subgiants \citep{bensby:13} and RC stars in the Baade's window \citep{hill:11}. We took a median metallicity of RC giants ([Fe/H]$=+0.16$) in \citet{hill:11} with a $0.11$~dex error. Additionally, the blue filled triangles with a dashed line represent the metallicity sensitivity of RC luminosity in \citet{nataf:13}, who adopted the distances and reddening values in the literature for 47~Tuc \citep[{\rm [Fe/H]}$=-0.76\pm0.04$;][]{koch:08} and NGC~6791 at the metallicity reported by \citet{brogaard:12}.

In Figure~\ref{fig:rclum}, our cluster data are shown by red filled squares, and are connected by a straight line to guide the eye. Absolute magnitudes from \citet{udalski:00}, \citet{groenewegen:08}, and that of the bulge indicate that the metallicity dependence of $M_I$ is mild below solar metallicity ($-0.5 \la {\rm [Fe/H]} \la 0$), but becomes steeper above solar. Previous $M_I$ measurements are qualitatively similar to the trend observed for the APOKASC-{\it Gaia} sample in $M_r$ (middle panel in Fig.~\ref{fig:mag}). The RC in 47~Tuc is fainter than the RC in M67, which may reflect lower RC masses in 47~Tuc. However, the above studies lack information on the masses of individual stars, which has a significant impact on absolute magnitudes, as revealed from the APOKASC-{\it Gaia} sample (Fig.~\ref{fig:mag}). Nonetheless, a collective trend observed from the previous studies, except the linear relation adopted by \citet{nataf:13}, is consistent with RC luminosities for the cluster sample.

\section{Photometric Mass Estimates of RC Giants}\label{sec:comp}

In this section, we proceeded with estimating RC masses by comparing the observed magnitudes of the cluster's RC with HB models. We employed a number of published HB models (\S~\ref{sec:models}), because models for highly evolved stars are subject to several systematic errors, despite continuing efforts to constrain input physics and parameters. We used RC giants in M67 to correct for zero-point offsets in colors and magnitudes predicted by each of these models (\S~\ref{sec:m67}). We employed the corrected models to read a mass gradation along the HB to estimate masses for the RC in NGC~6791 and field RC giants in the APOKASC-{\it Gaia} sample (\S~\ref{sec:n6791}). The advantage of studying cluster systems is that we know a priori the cluster's metallicity and age well, so the integrated amount of mass loss on the RGB directly follows from a comparison of the RC mass with its initial mass. One of the principal results of this paper is the striking degree of disagreement between published theoretical models of core He-burning stars at the high-metallicity end.

\subsection{ZAHB Models}\label{sec:models}

\begin{figure}
\epsscale{1.2}
\plotone{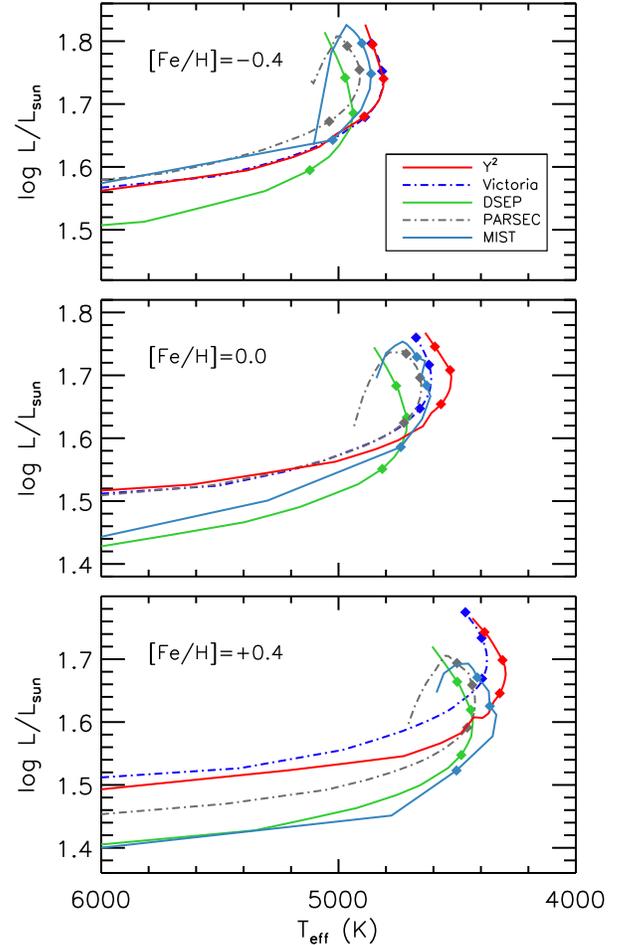}
\caption{Comparisons of theoretical ZAHB models at [Fe/H]$=-0.4$ (top), solar (middle), and [Fe/H]=$+0.4$ (bottom). Along each grid line, $0.7\ M_\odot$, $1.0\ M_\odot$, and $1.3\ M_\odot$ models are marked by filled diamonds.\label{fig:models}} \end{figure}

Figure~\ref{fig:models} displays the ZAHB models employed in this study. These models were taken from evolutionary tracks generated using (1) the Dartmouth Stellar Evolution Program \citep[DSEP;][]{dotter:08} at [Fe/H]$=-0.5$, $0.0$, $0.3$, and $0.5$; (2) MIST \citep{choi:16,dotter:16} at $-0.5 \leq {\rm [Fe/H]} \leq 0.5$ in $0.25$~dex increments; (3) PARSEC \citep{bressan:12} at $-0.6 \leq {\rm [Fe/H]} \leq 0.7$ in $\sim0.1$--$0.2$~dex increments; (4) Victoria-Regina \citep[Victoria;][]{vandenberg:06} at $-0.6 \leq {\rm [Fe/H]} \leq 0.5$ in approximately $0.1$~dex increments; and (5) Yonsei-Yale \citep[Y$^2$;][]{yi:03,chung:13} at $-0.5 \leq {\rm [Fe/H]} \leq 0.5$ in $0.1$~dex increments. For each set of models, we made a finer-mass grid with a step size of $\Delta M = 0.01\ M_\odot$ using a quadratic interpolation. All of the models assume scaled solar abundances. The only exception was the DSEP models at [Fe/H]$=-0.5$, for which we assumed [$\alpha$/Fe]$=0.2$. However, the effect of [$\alpha$/Fe] is relatively minor: models with $\alpha$ enhancement by [$\alpha$/Fe]$=0.2$ are similar to solar-scaled models with a $\Delta$[Fe/H]$=0.1$ increment.

In order to compare theoretical models with observations, the temperatures and luminosities predicted in the models were converted into observable colors and magnitudes using color-$\teff$ relations and bolometric corrections. We utilized the color transformations found in \citet{girardi:02,girardi:04}, which were largely based on the ATLAS9 spectral library \citep[][hereafter ATLAS9]{castelli:97,bessell:98} over the parameter space covered by RC giants in this work. In addition, we  took theoretical HB tracks from the DSEP with theoretical colors and magnitudes based on PHOENIX model spectra \citep[][hereafter PHOENIX]{hauschildt:99}. Similarly, we made use of the ZAHB models from Y$^2$ with synthetic colors and magnitudes from the BaSeL spectral libraries \citep[][hereafter BaSeL]{westera:02}. We converted model magnitudes in the AB system into the native SDSS system using AB corrections found in \citet{eisenstein:06}: $i_{\rm AB} = i_{\rm SDSS}-0.015$. No corrections are required in the $g$ and $r$ passbands. Throughout the paper, models are denoted as ``interior+atmosphere'' models, such as DSEP+ATLAS9.

Figures~\ref{fig:comp67} and \ref{fig:comp6791a} show comparisons of the observed locations of the RC in M67 and NGC~6791 with six different combinations of ZAHB models at [Fe/H]$=0.0$ and $=+0.4$, respectively. The ZAHB lines from different models do not agree with each other, which can lead to an inaccurate mass estimate of RC giants from CMDs. In particular, DSEP models consistently predict fainter magnitudes by $\Delta M_r \sim 0.2$~mag than the other models, primarily due to their systematically lower luminosities (Figure~\ref{fig:models}). At a given mass, there are also differences in the predicted colors of the order of $\sim0.1$--$0.2$~mag. The differences are amplified at supersolar metallicity.

\begin{deluxetable*}{lcccccc}
\tablewidth{0pt}
\tablecaption{Initial Helium and Metal Abundances of HB Models Employed in This Work\label{tab:yz}}
\tablehead{
  \colhead{} &
  \colhead{Primordial} &
  \colhead{Helium} &
  \multicolumn{4}{c}{[Fe/H]} \\
  \cline{4-7}
  \colhead{Model} &
  \colhead{Helium Abundance} &
  \colhead{Enrichment} &
  \multicolumn{2}{c}{$0.00$} &
  \multicolumn{2}{c}{$+0.37$} \\
  \cline{4-5}
  \cline{6-7}
  \colhead{Name} &
  \colhead{$(Y_p)$} &
  \colhead{$(\Delta Y/\Delta Z)$} &
  \colhead{$Y$} &
  \colhead{$Z$} &
  \colhead{$Y$} &
  \colhead{$Z$}
}
\startdata
DSEP     & $0.245$   & $1.54$  & $0.274$ & $0.019$ & $0.308$ & $0.041$ \\
MIST     & $0.249$   & $1.5$   & $0.270$ & $0.014$ & $0.296$ & $0.031$ \\
PARSEC   & $0.2485$  & $1.78$  & $0.276$ & $0.015$ & $0.306$ & $0.032$ \\
Victoria & $0.23544$ & $2.2$   & $0.277$ & $0.019$ & $0.323$ & $0.040$ \\
Y$^2$    & $0.23$    & $2.0$   & $0.266$ & $0.018$ & $0.308$ & $0.039$ 
\enddata
\end{deluxetable*}

The observed difference among these models indicates that various assumptions and detailed calculations in the models affect the ZAHB location and shape. Among them, the assumed helium abundance is of particular importance, because the ZAHB luminosity is a sensitive function of the helium abundance and because its effect is enhanced in metal-rich stars by the CNO cycle in the hydrogen-burning shell \citep[e.g.,][]{sweigart:76,valcarce:12}. A common practice is to set an initial value of a helium mass fraction ($Y$) based on an approximate relation with a mass fraction of heavy elements ($Z$), through $Y=Y_p+(\Delta Y/\Delta Z)Z$, where $Y_p$ is a primordial helium abundance and $\Delta Y/\Delta Z$ represents a helium enrichment parameter. The second and third columns in Table~\ref{tab:yz} list the $Y_p$ and $\Delta Y/\Delta Z$, respectively, adopted by each model. The initial helium abundances computed from these parameters are listed at [Fe/H]$=0.0$ and $+0.37$. The heavy-element fraction ($Z$) in each model is different from one another due to differences in the adopted solar abundance [\citet{grevesse:98} vs.\ \citet{asplund:09}].

From a large grid of Padova models in $Y$ and $Z$ \citep{bertelli:08}, we found an $\sim\pm0.1$~dex change in luminosity from $\Delta Y=\pm0.05$ at a given $Z$ without a significant shift in $\teff$. However, the differences in the helium fraction are not sufficiently large enough to explain the observed scatter of ZAHB models. This is especially true at supersolar metallicities. All of the models employed in this work, except Victoria, show similar $Y$ values at [Fe/H]$=0.37$ (Table~\ref{tab:yz}) that are in full agreement with $Y=0.30\pm0.01$ from the eclipsing binary study in NGC~6791 \citep{brogaard:12} and $Y=0.297\pm0.003$ based on detailed analysis of asteroseismic observations \citep{mckeever:19}. This indicates that there should be other sources of systematic errors embedded in the stellar interior models that are large enough to induce significant effects on predicted luminosities and temperatures of HB stars. In the Appendix, we present additional comparisons of models employed in this work.

\subsection{Zero-point Corrections of ZAHB Models Using RC Stars in M67}\label{sec:m67}

The diversity of interior models on the theoretical Hertzsprung-Russell Diagram (Figure~\ref{fig:models}) indicates the extent of the parameter space that is poorly constrained by our current knowledge of stellar structures. This is particularly true for luminosity, which is a robustly predicted quantity in stellar interior models. Its scatter reflects different treatment of physical processes (e.g., convective overshoot and microscopic diffusion) and/or assumptions on physical parameters (e.g., mixing-length parameter and nuclear reaction rate). Differences in numerical methods might also play a role. It is beyond the scope of this work to dissect interior models and identify the origin of such differences. Instead, we take a heuristic approach below, in which we derive RC masses using models with zero-point shifts in colors and magnitudes to match those of the RC in M67.

Direct comparisons of RC giants in M67 with ZAHB models are shown in Figure~\ref{fig:comp67}, for each set of combinations of interior and atmospheric models at [Fe/H]$=0$. Each of the theoretical ZAHB lines is bifurcated into two branches by luminosity, which is useful for constraining the range of the cluster's RC mass. The inflection point at $\gr \sim 0.9$ corresponds to $0.8\ M_\odot$, and more (less) massive stars are brighter (fainter) than this inflection point.  At the observed color of RC giants, the gap between the low- and high-luminosity branches is about $0.4$~mag in $r$ or $V$.

As expected from Figure~\ref{fig:models}, however, different models show large scatter in the predicted colors and magnitudes, even at solar metallicity. The scatter is large enough to prevent a meaningful constraint on stellar mass; if masses are inferred from $M_V$, they range from $\sim1.3$ to $\sim1.7\ M_\odot$, depending on which model is employed in the photometric mass estimate. The problem is even worse at supersolar metallicities (Figure~\ref{fig:comp6791a}).

In order to derive RC masses, we corrected models on an empirical basis to match observed data from a well-studied system. We did this by adjusting the colors and magnitudes of models using RC giants in M67. This implies that all of the systematic errors in the models are absorbed into errors in the bolometric magnitude scales. To first order, this should provide sufficiently accurate mass estimates at or near solar metallicity. Below we applied these corrected models to subsolar- and supersolar-metallicity stars (including NGC~6791), hoping that color and magnitude offsets are effectively canceled out in the extended parameter space.

\begin{figure*}
\epsscale{0.95}
\plotone{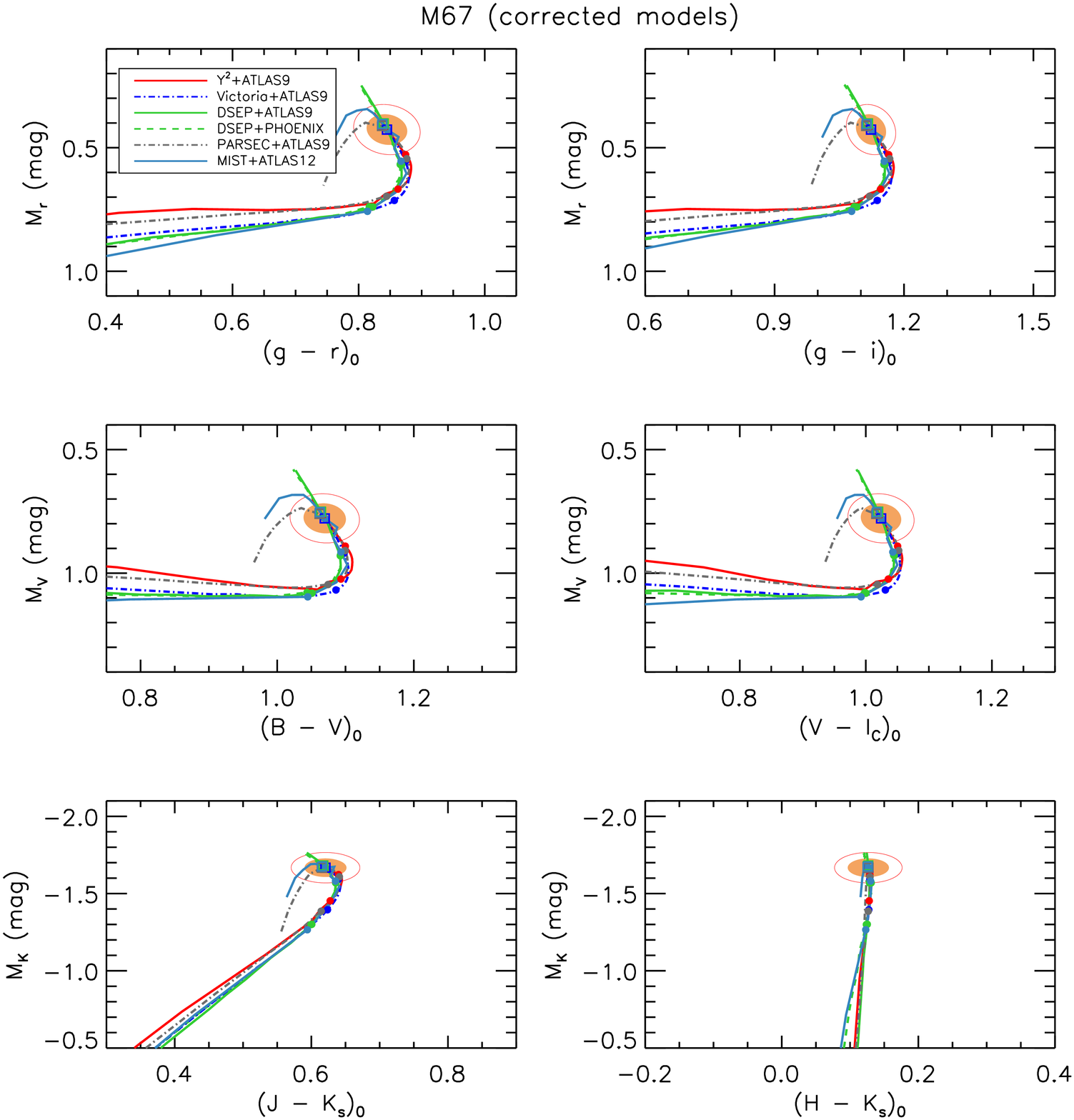}
\caption{Same as Figure~\ref{fig:comp67} but displaying corrected ZAHB models.\label{fig:comp67b}} \end{figure*}

\begin{deluxetable*}{lccccccc}
\tablewidth{0pt}
\tablecaption{Zero-point Corrections on ZAHB Models\label{tab:zp}}
\tablehead{
  \colhead{} &
  \multicolumn{7}{c}{Models} \\
  \cline{2-8}
  \colhead{Magnitudes/} &
  \colhead{DSEP} &
  \colhead{DSEP} &
  \colhead{MIST} &
  \colhead{PARSEC} &
  \colhead{Victoria} &
  \colhead{Y$^2$} &
  \colhead{Y$^2$} \\
  \colhead{Colors} &
  \colhead{+ATLAS9} &
  \colhead{+PHOENIX} &
  \colhead{+ATLAS9} &
  \colhead{+ATLAS9} &
  \colhead{+ATLAS9} &
  \colhead{+ATLAS9} &
  \colhead{+BaSeL} \\
  \colhead{} &
  \colhead{(mag)} &
  \colhead{(mag)} &
  \colhead{(mag)} &
  \colhead{(mag)} &
  \colhead{(mag)} &
  \colhead{(mag)} &
  \colhead{(mag)}
}
\startdata
$M(r)$            & $-0.185$ & $-0.147$ & $-0.075$ & $-0.072$ & $-0.025$ & $-0.088$ & $-0.003$ \\
$(g\, -\, r)_0$   & $-0.005$ & $-0.021$ & $-0.034$ & $-0.026$ & $-0.048$ & $-0.087$ & $-0.049$ \\
$(g\, -\, i)_0$   & $+0.011$ & $-0.025$ & $-0.034$ & $-0.019$ & $-0.049$ & $-0.105$ & $-0.073$ \\
\cline{1-8}                                                                              
$M(V)$            & $-0.118$ & $-0.135$ & $-0.020$ & $-0.015$ & $+0.024$ & $-0.056$ & $-0.010$ \\
$(B\, -\, V)_0$   & $-0.010$ & $-0.040$ & $-0.023$ & $-0.031$ & $-0.051$ & $-0.084$ & $-0.036$ \\
$(V\, -\, I_C)_0$ & $-0.029$ & $-0.032$ & $-0.025$ & $-0.049$ & $-0.070$ & $-0.108$ & $+0.054$ \\
\cline{1-8}                                                                              
$M(K_s)$          & $-0.131$ & $-0.183$ & $+0.023$ & $+0.023$ & $+0.113$ & $+0.129$ & $-0.055$ \\
$(J\, -\, K_s)_0$ & $+0.029$ & $+0.019$ & $-0.014$ & $+0.015$ & $+0.000$ & $-0.029$ & $+0.247$ \\
$(H\, -\, K_s)_0$ & $+0.043$ & $+0.045$ & $+0.010$ & $+0.042$ & $+0.040$ & $+0.037$ & $+0.189$
\enddata
\end{deluxetable*}

We assumed that the current mean RC mass in M67 is $1.3\ M_\odot$ and computed zero-point corrections in colors and magnitudes for each model set. Table~\ref{tab:zp} lists photometric zero-point offsets that have been added to match each ZAHB model to the position of the RC in M67. Typical color offsets amount to a few hundredth of a magnitude, or an $\sim50$--$100$~K shift in terms of $\teff$, although some models require larger shifts. Figure~\ref{fig:comp67b} shows corrected ZAHB models at solar metallicity. Since more massive RC giants are brighter, our working hypothesis on the minimal mass loss in M67 effectively sets an upper limit in the RC mass or a lower limit in the amount of RGB mass loss. In other words, if we were adopting a lower RC mass in M67 (more mass loss along RGB), our photometric RC mass would become subsequently smaller.

\begin{figure*}
\epsscale{0.95}
\plotone{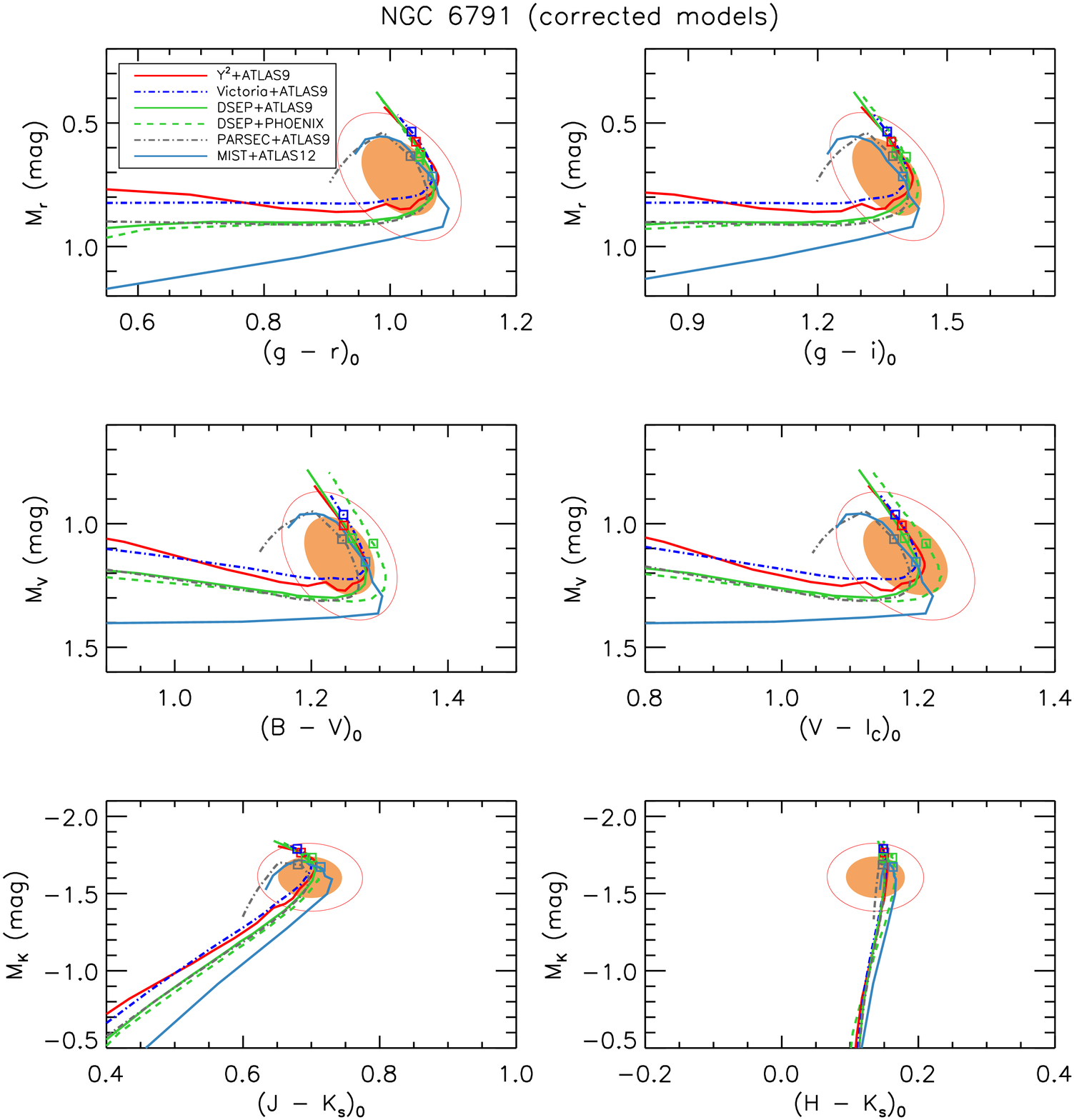}
\caption{Same as Figure~\ref{fig:comp6791a} but displaying corrected ZAHB models (see text). Error ellipses include additional error contributions from those of RC giants in M67.\label{fig:comp6791b}} \end{figure*}

Figure~\ref{fig:comp6791b} shows comparisons of the RC in NGC~6791 with corrected ZAHB models at [Fe/H]$=0.37$. The same amounts of shifts in colors and magnitudes as in the solar-metallicity case have been applied to these models. An error ellipse includes errors in the RC location in M67. While the zero-point adjustment certainly improves consistency among models (compare with Figure~\ref{fig:comp6791a}), there still remain relatively large differences at supersolar metallicity. This indicates that the above first-order corrections based on M67 are not perfect, suggesting nonlinear effects of stellar parameters at large metallicities. Below, we show that the RC masses in NGC~6791 obtained using these corrected models also differ by the amounts that are too large to be explained by internal errors.

\subsection{Photometric Mass of RC Giants}\label{sec:n6791}

\subsubsection{NGC~6791}

In contrast to M67, RC stars in NGC~6791 are located in or near the faint branch of the models (Figure~\ref{fig:comp6791b}). The relatively lower luminosity of the RC indicates that their masses are smaller. In order to derive the best-matching photometric mass and its confidence interval for each model set, we computed a marginal likelihood of RC mass, $\mathcal{M}$, for an observed data set in all nine bandpasses, $X = \{X_i\} = \{g, r, i, B, V, I_C, J, H, K_s\}$, as follows:
\begin{eqnarray}
\mathcal{L} (\mathcal{M}; X) =  \prod_{i=1}^{N} {\rm prob} (X_i | \mathcal{M}) \label{eq:likelihood}
\\
\propto \prod_{i=1}^{N} \int {\rm prob} (X_i | \vec{\alpha}, \mathcal{M})\, {\rm prob} (\vec{\alpha} | \mathcal{M}) {\rm d}\vec{\alpha} \label{eq:likelihood}
\end{eqnarray}
,where $\vec{\alpha}$ represents the joint parameters $\{\dmn$, [Fe/H], $\ebv$, $R_V$, $R_{VI}$, $R_{VK}$, $R_{JK}$, $R_{HK}$, $A_g/A_V$, $A_r/A_V$, $A_i/A_V\}$ that are marginalized out. We assumed a normal distribution of errors in the prior probability functions of these nuisance parameters. Here we computed a probability density function prob($X_i$) by taking a minimum $\chi^2$ of the observed magnitude from models with a step size of $0.01\ M_\odot$.

\begin{figure}
\epsscale{1.2}
\plotone{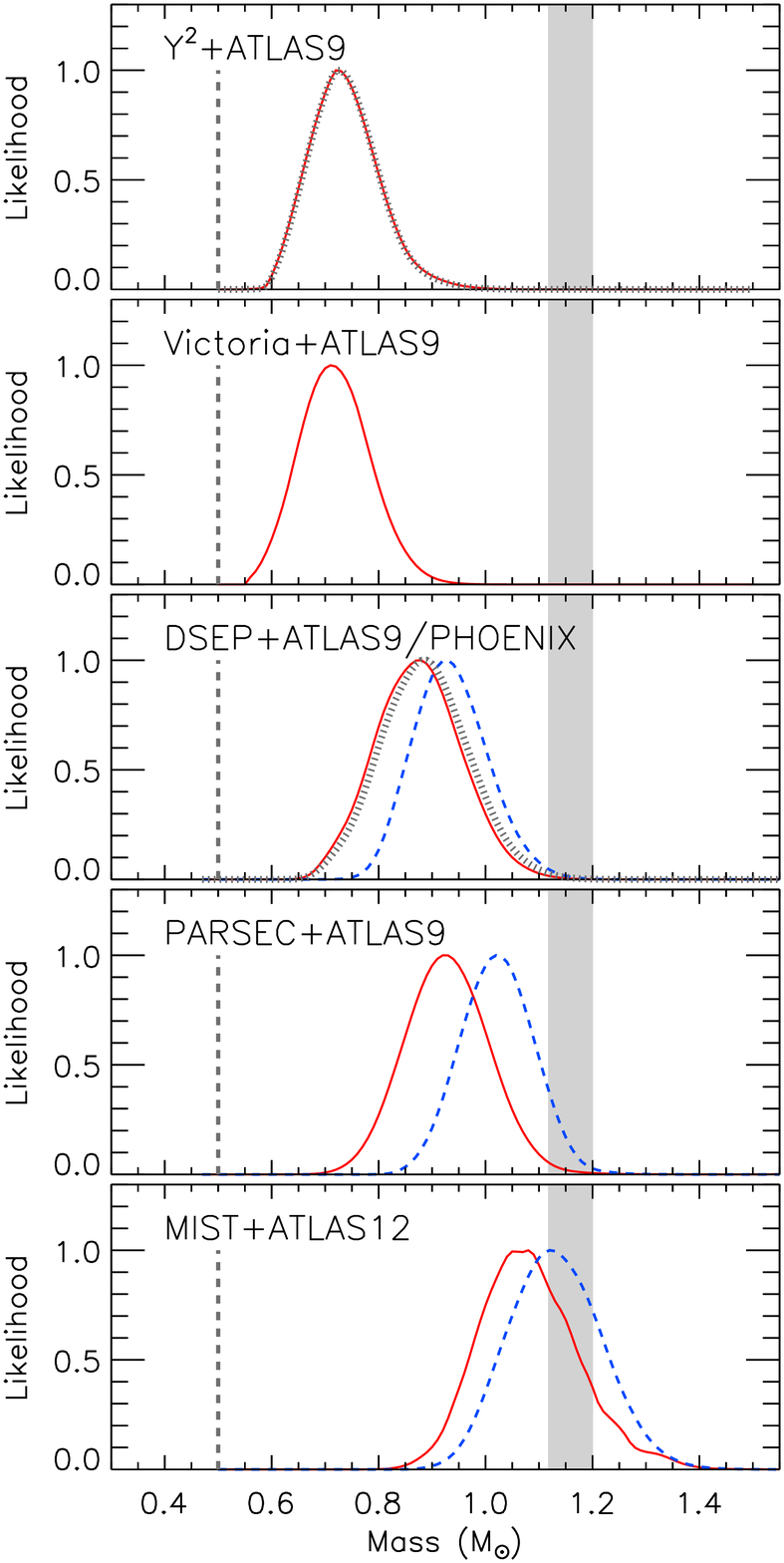}
\caption{Normalized likelihood functions of the mean RC mass in NGC~6791 based on different ZAHB models with ATLAS9 color transformations (red curve). Additional results from Y$^2$+BaSeL and DSEP+PHOENIX are shown by gray dotted curves. Similarly, the blue dashed curves in the bottom three panels represent the likelihood of the RC mass after taking evolutionary effects into account. All of the likelihood distributions are based on the corrected models (see text). In all panels, the gray shaded column represents a $\pm1\sigma$ range of the asteroseismic mass of RC giants. The vertical dashed lines mark the masses of EHB stars in the cluster.\label{fig:likelihood}} \end{figure}

\begin{deluxetable}{lccc}
\tablewidth{0pt}
\tablecaption{Photometric Mass of RC Giants in NGC~6791\label{tab:mass}}
\tablehead{
  \colhead{} &
  \colhead{} &
  \multicolumn{2}{c}{Confidence Intervals} \\
  \cline{3-4}
  \colhead{{\bf Corrected Models}\tablenotemark{a}} &
  \colhead{Mode} &
  \colhead{$68.3\%$} &
  \colhead{$95.5\%$} \\
  \colhead{} &
  \colhead{$(M_\odot)$} &
  \colhead{$(M_\odot)$} &
  \colhead{$(M_\odot)$}
}
\startdata
DSEP(ZAHB)+ATLAS9     & $0.88$ & $_{0.79}^{0.95}$ & $_{0.71}^{1.01}$ \\
DSEP(evol)+ATLAS9     & $0.93$ & $_{0.85}^{1.00}$ & $_{0.80}^{1.05}$ \\
DSEP(ZAHB)+PHOENIX    & $0.89$ & $_{0.80}^{0.96}$ & $_{0.72}^{1.03}$ \\
MIST(ZAHB)+ATLAS9     & $1.08$ & $_{0.98}^{1.17}$ & $_{0.91}^{1.25}$ \\
MIST(evol)+ATLAS9     & $1.12$ & $_{1.03}^{1.21}$ & $_{0.96}^{1.28}$ \\
PARSEC(ZAHB)+ATLAS9   & $0.92$ & $_{0.84}^{0.99}$ & $_{0.77}^{1.06}$ \\
PARSEC(evol)+ATLAS9   & $1.02$ & $_{0.94}^{1.08}$ & $_{0.88}^{1.13}$ \\
Victoria(ZAHB)+ATLAS9 & $0.71$ & $_{0.64}^{0.77}$ & $_{0.59}^{0.83}$ \\
Y$^2$(ZAHB)+ATLAS9    & $0.72$ & $_{0.66}^{0.79}$ & $_{0.61}^{0.84}$ \\
Y$^2$(ZAHB)+BaSeL     & $0.73$ & $_{0.66}^{0.79}$ & $_{0.62}^{0.84}$ \\
\hline
Weighted mean\tablenotemark{b} & \multicolumn{3}{c}{$0.88\pm0.16$}
\enddata
\tablenotetext{a}{evol: models connecting the most likely locations of HB stars, which take into account evolutionary effects.}
\tablenotetext{b}{Weighted mean RC mass from the ZAHB+ATLAS9 models, after adding a systematic offset from the solutions based on the `evol' models (see text).}
\end{deluxetable}

Figure~\ref{fig:likelihood} shows the likelihood distributions of the mean RC mass in NGC~6791 for each combination of ZAHB models employed in this work. Each curve is normalized with respect to the maximum likelihood value. Table~\ref{tab:mass} lists modes of RC mass and their $68.3\%$ and $95.5\%$ confidence intervals. The estimated mass ranges from $\sim0.7$ to $\sim1.1\ M_\odot$. Among these models, the MIST models yield the largest RC mass, while Victoria and Y$^2$ provide the lowest masses. This is closely related to the helium abundance of the metal-rich models (Table~\ref{tab:yz}). Because HB models with lower helium abundances are fainter, photometric masses from these models are systematically larger. MIST models adopt a small $\Delta Y/\Delta Z$, resulting in a relatively low helium abundance at [Fe/H]$=+0.37$.

Stars with $0.7\ M_\odot \la M_* \la 1.5\ M_\odot$ generally evolve toward lower luminosities when they leave the ZAHB \citep[e.g., see Figure~3 in][]{girardi:16}, and photometric mass estimates become larger. We repeated the above exercise by taking an HB with the most likely distribution from individual stellar tracks. The results are shown by a blue dashed curve in Figure~\ref{fig:likelihood} for the DSEP, MIST, and PARSEC models, and their photometric masses are listed in Table~\ref{tab:mass} (denoted as `evol'). The difference in mass from the ZAHB case is about $0.05$-$0.10\ M_\odot$.

We also checked how our results depend on different choices of color-$\teff$ relations and bolometric corrections. As shown in Table~\ref{tab:mass}, photometric masses based on other atmospheric models (BaSeL and PHOENIX) are almost the same as those from ATLAS9 ($\Delta M \sim 0.01\ M_\odot$). At solar metallicity, the synthetic magnitudes predicted by these models differ by $0.04$--$0.09$~mag in $r$ and $0.02$--$0.05$~mag in $V$ for $1.3\ M_\odot$ models (Table~\ref{tab:zp}). Additionally, we compared the ATLAS9-based bolometric corrections in \citet{girardi:02,girardi:04} with model fluxes in \citet{casagrande:14} from MARCS \citep{marcs}, and found similar differences. These systematic offsets in bolometric magnitudes between different models remain almost unchanged at supersolar metallicities. As a result, photometric masses become almost independent of adopted color-$\teff$ relations and bolometric corrections, because our zero-point corrections of HB models based on M67 essentially remove any model-to-model differences.

The weighted mean value of RC mass in NGC~6791 is $0.83\pm0.15\ M_\odot$ from all of the ZAHB models with ATLAS9. The error represents a weighted standard deviation of the five solutions and is about a factor of $2$ larger than a propagated error, which suggests a strong model dependence of the photometric mass estimate. The systematic error component from a difference between ZAHB and `evolved' models makes photometric mass estimates larger. Therefore, we added $0.05\ M_\odot$ to the weighted mean value from the above comparisons and also added the same value in quadrature to the error budget. The effects of different choices of atmosphere models are negligible. Based on these considerations, our photometric RC mass becomes $0.88\pm0.16\ M_\odot$.

There have also been a couple of attempts to constrain RC masses photometrically. For example, \citet{carraro:96} found $0.9\ M_\odot$ for an average mass of RC giants in NGC~6791, based on their global isochrone fitting on a CMD [$\ebv=0.15$ and apparent distance modulus $\dmnv=13.50$]. On the other hand, \citet{brogaard:12} showed that the cluster RC giants are well matched to the updated Victoria-Regina models\footnote{The Victoria models utilized in our study \citep{vandenberg:06} are different from those used in \citet{brogaard:12} in that the latter models were computed based on more up-to-date physics, including microscopic diffusion of helium (D.\ VandenBerg 2019, private communication). See the Appendix for more information.}, if one assumes a small RGB mass loss in \citet{miglio:12}, suggesting a larger RC mass ($\sim 1$--$1.15\ M_\odot$), although their assumed distance and reddening were similar to those used in \citet{carraro:96}. Since the helium and metal abundances assumed in these two studies are also similar to each other, the difference in mass can be traced back to systematic differences between stellar models employed by each study. The RC mass in \citeauthor{brogaard:12} is also larger than our estimated value from the Victoria models ($0.71\pm0.07\ M_\odot$). A part of the difference can be explained by small differences in the adopted cluster distance and reddening, but it is mainly due to input stellar models (see Figure~\ref{fig:model5} in the Appendix).

\subsubsection{Field RC Giants}\label{sec:field}

\begin{figure*}
\epsscale{1.0}
\plottwo{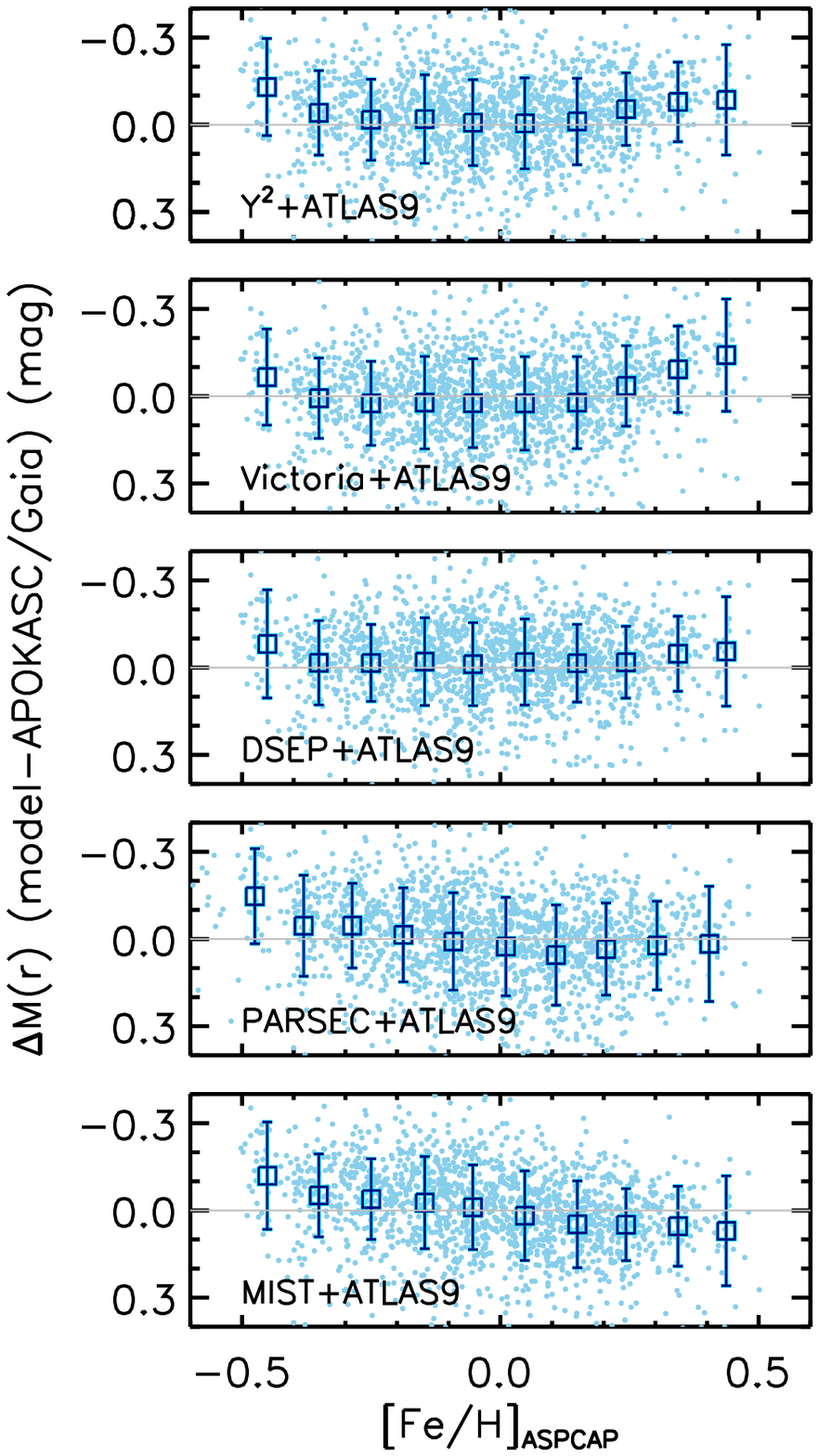}{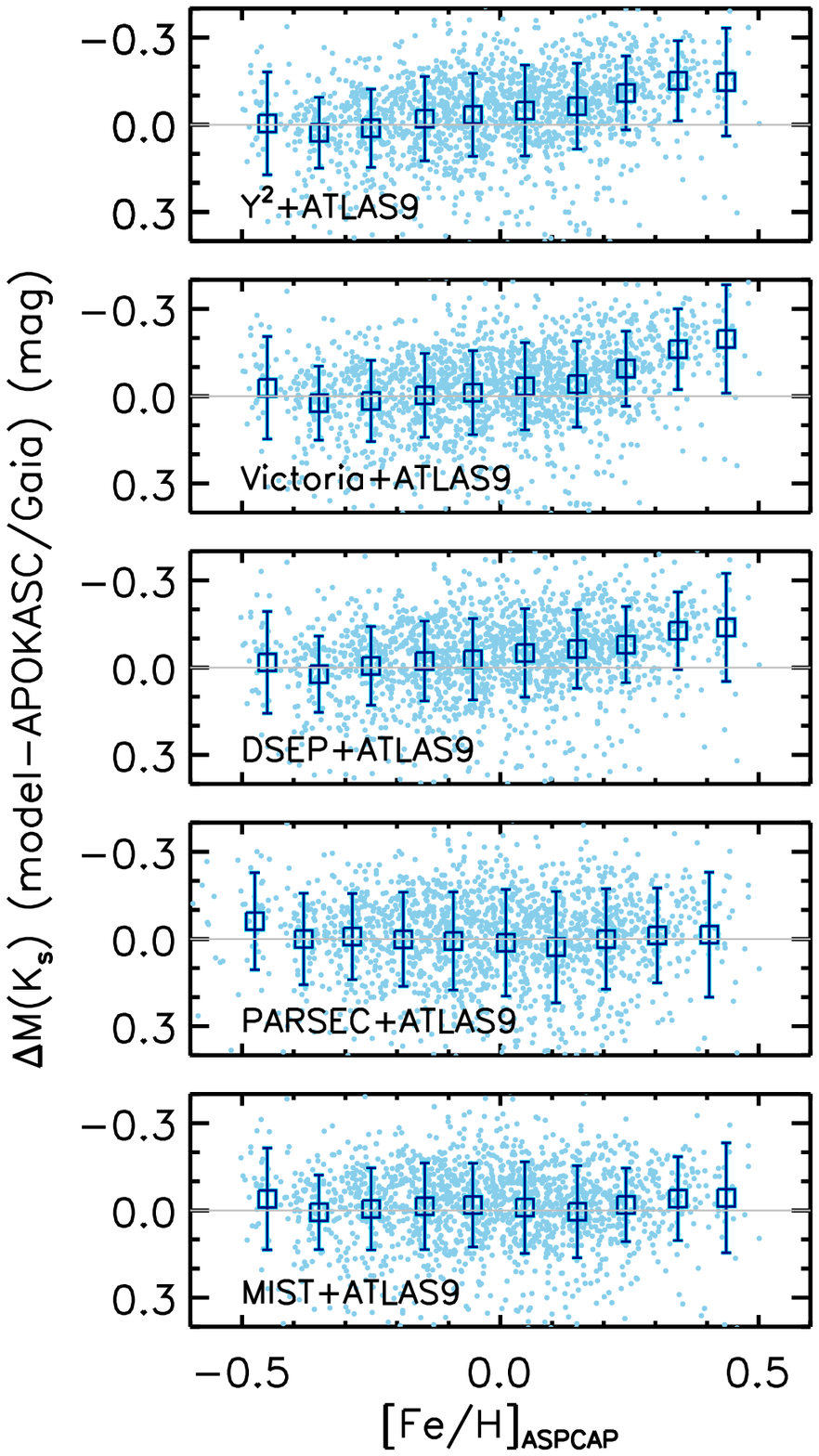}
\caption{Differences between the $M_r$ (left) and $M_K$ (right) of the APOKASC-{\it Gaia} sample and those inferred from a set of theoretical ZAHB models. Squares represent an average difference in each metallicity bin, where the error bars indicate a standard deviation of the magnitude difference. A small number of outliers are excluded in the above plots.\label{fig:mr}} \end{figure*}

In addition to NGC~6791, we repeated the above photometric mass estimation for field RC giants in the APOKASC-{\it Gaia} sample. Because these stars lack homogeneous and accurate photometry in $BVI_C$, we only included $griJHK_s$ in the likelihood estimation in Equations~(1) and (2). Figure~\ref{fig:mr} shows comparisons between the absolute magnitudes ($M_r$ and $M_K$) of the APOKASC-{\it Gaia} sample and those predicted by each set of ZAHB models. We computed the absolute magnitudes in the models by taking the asteroseismic masses and spectroscopic metallicities. We adopted extinctions in the APOKASC catalog and limited the comparison to those with $\sigma_\pi / \pi < 0.03$ and $d_\odot < 5$~kpc. In each panel, we computed the average masses for a given [Fe/H] bin, which are shown by open squares. Error bars indicate a standard deviation of the differences in each bin.

At solar metallicity, values of $M_r$ from all five of the models shown in Figure~\ref{fig:mr} are consistent with estimates in the APOKASC-{\it Gaia} catalog. This is a natural consequence of our adjustment of models to match the location of the RC in M67. Nonetheless, the difference is not exactly zero ($ | \Delta M_r | \la 0.02$~mag and $| \Delta M_K | \la 0.05$~mag). This is probably because most of the APOKASC-{\it Gaia} sample included in Figure~\ref{fig:mr} have lower masses than the mass of the RC in M67 ($\sim1.3\ M_\odot$), suggesting a scale error in mass either in the models or in the asteroseismic scaling relations.

All of the ZAHB models, except PARSEC+ATLAS9, exhibit a metallicity-dependent departure from the observed absolute magnitudes of stars. At the metallicity of NGC~6791, Y$^2$+ATLAS9, Victoria+ATLAS9, and DSEP+ATLAS9 predict a brighter HB by $\Delta M_r \approx 0.05$--$0.10$~mag than the APOKASC-{\it Gaia} sample, while the MIST+ATLAS9 models become fainter. Comparisons of $M_K$ with the former models show a strong tilt with [Fe/H], and the difference reaches $| \Delta M_K | \approx 0.15$--$0.20$ at the metallicity of NGC~6791. 

The mismatches seen in Figure~\ref{fig:mr} do not necessarily indicate problems in a specific set of ZAHB models employed in this study. Asteroseismic masses are used as an input to obtain the absolute magnitudes of the models. Therefore, if there is a scale error in mass between asteroseismology and HB models at higher metallicities, a systematically brighter or fainter HB can be derived from the models. On the other hand, errors in foreground extinctions or distances are unlikely to cause the observed departures, because these parameters in the sample are not correlated with metallicity.

\begin{figure}
\epsscale{1.2}
\plotone{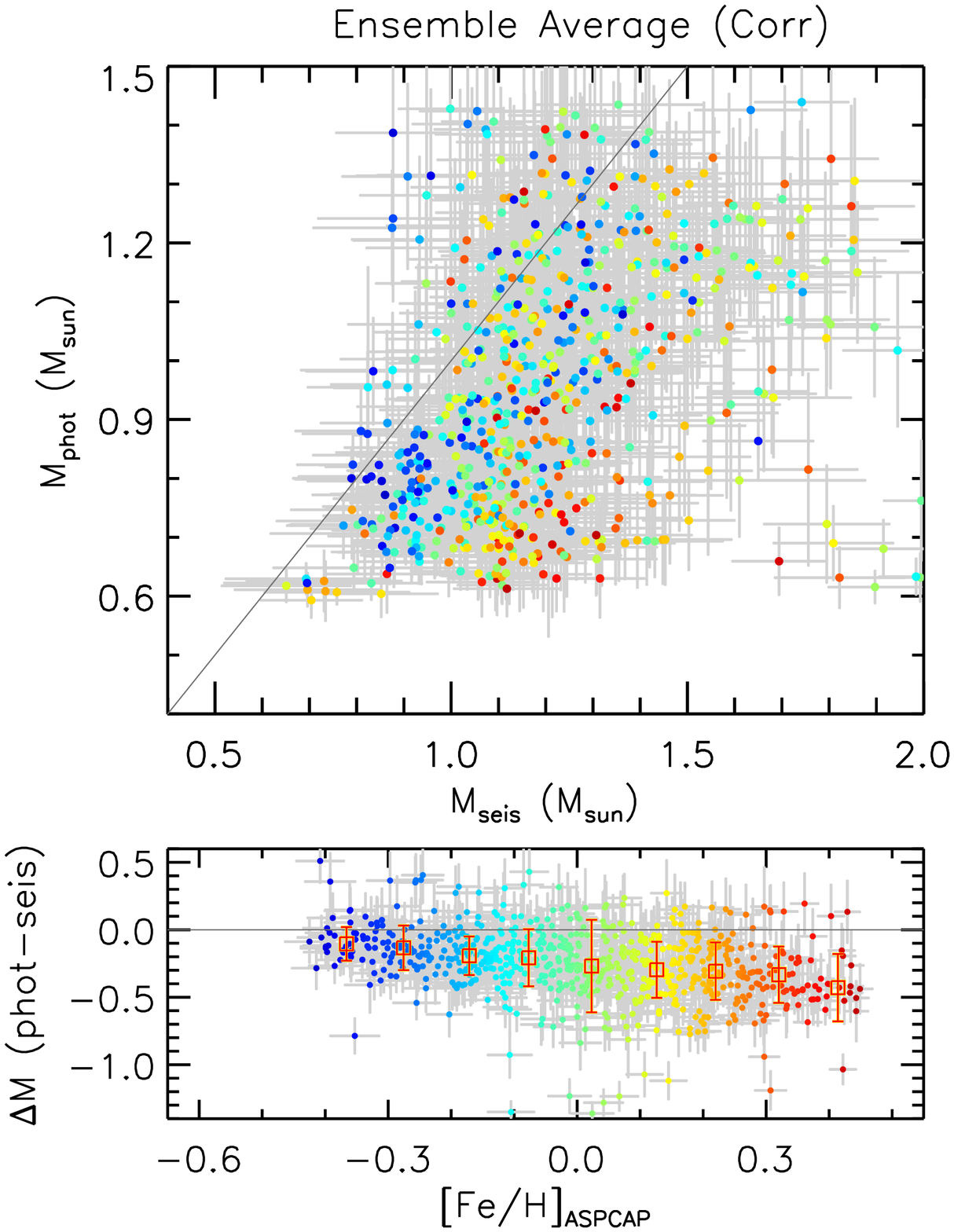}
\caption{Comparison between asteroseismic and photometric masses of RC giants in the APOKASC-{\it Gaia} catalog. Photometric masses represent ensemble averages from five ZAHB model sets (DSEP, MIST, PARSEC, Victoria, and $Y^2$ with ATLAS9 colors) and are shown if there are at least three valid mass estimates (see text). Colors represent a metallicity of each star. In the bottom panel, binned average differences are shown by squares, along with a standard deviation by an error bar.\label{fig:ensemble}} \end{figure}

In Figure~\ref{fig:ensemble}, we compare photometric masses with asteroseismic estimates for the APOKASC-{\it Gaia} sample. We took an ensemble average of photometric masses from the above five ZAHB models with ATLAS9. We imposed $\sigma_\pi/\pi < 0.06$ and $d_\odot < 2$~kpc to minimize the effect of extinction while maintaining a useful number of stars for comparison. We computed weighted mean averages if there are valid photometric mass estimates ($\sigma_M < 0.5\ M_\odot$ and a total $\chi^2$ from the best-fitting model $\chi_{\rm tot}^2 < 5$) from at least three HB models. In this ensemble average, we only included stars with $M_{\rm phot} < 1.5\ M_\odot$, because photometric estimates become progressively more uncertain and partially degenerate with higher-mass solutions (see MIST models in Figure~\ref{fig:comp6791a}). Error bars indicate a quadrature sum of errors from both random (propagated errors from individual models) and systematic (model-to-model dispersion) components.

As seen in Figure~\ref{fig:ensemble}, average photometric masses are generally smaller than asteroseismic masses, where the difference at solar metallicity is almost $0.25\ M_\odot$. The observed trend depends strongly on metallicity and increases to $\sim0.4\ M_\odot$ at [Fe/H]$=+0.4$. The sense and size of the difference are about the same as those found for the RC in NGC~6791. The strong metallicity dependence of the difference is mostly driven by systematically brighter HBs in Y$^2$+ATLAS9 and Victoria+ATLAS9 (see Figure~\ref{fig:mr}).

\begin{figure*}
\epsscale{1.0}
\plottwo{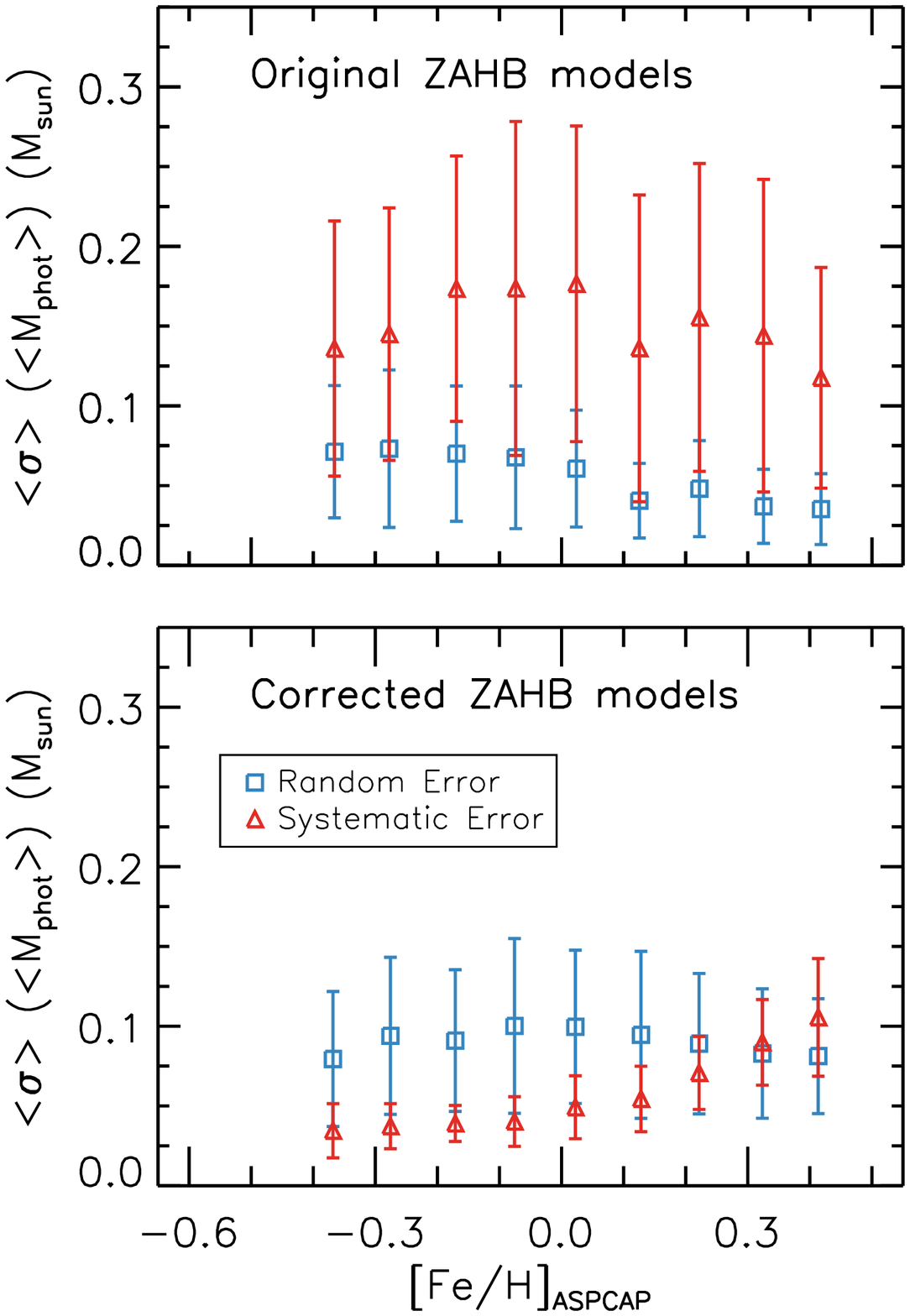}{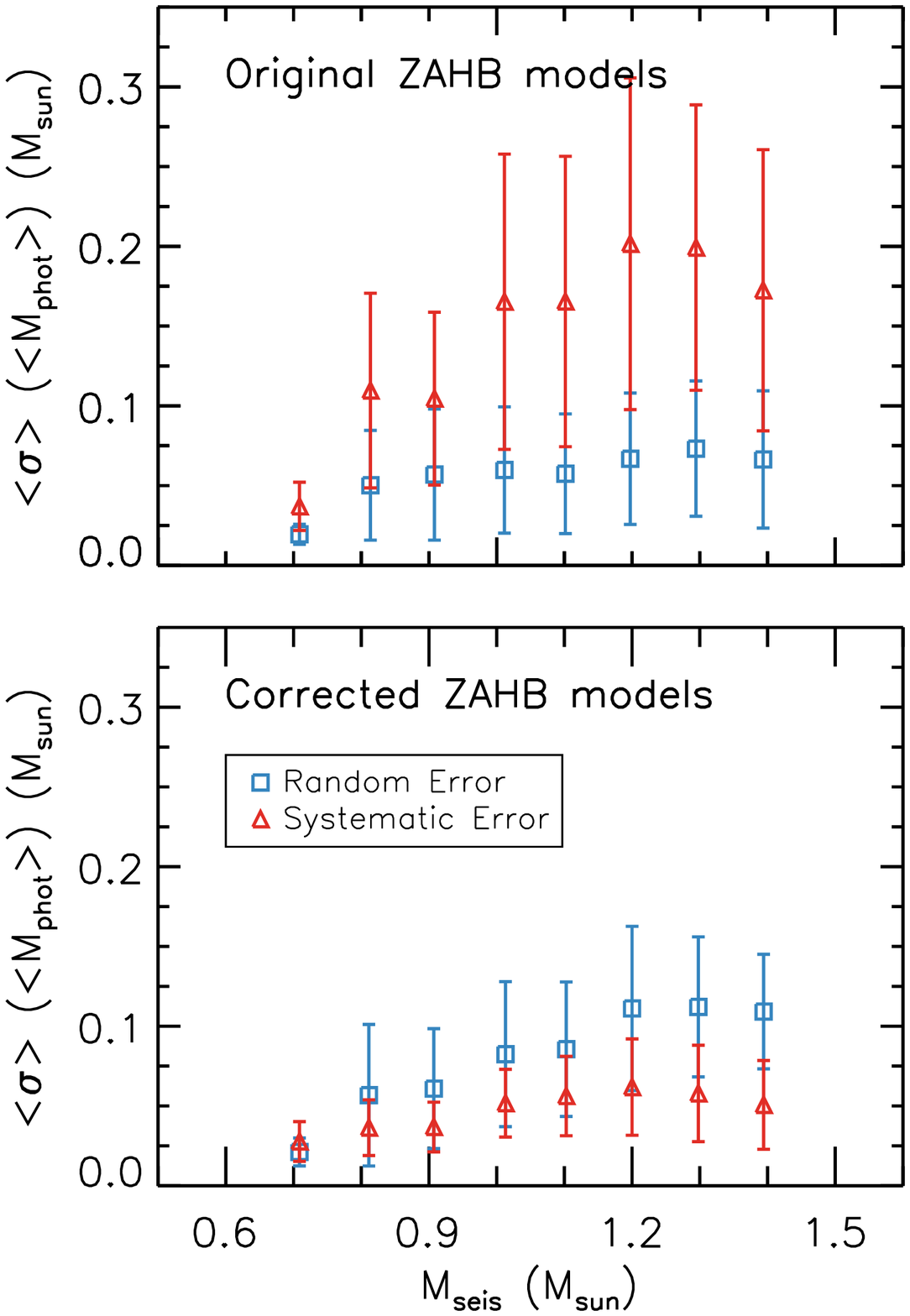}
\caption{Error distributions of ensemble mass averages from the original (top) and corrected (bottom) ZAHB models, as a function of [Fe/H] (left) and asteroseismic mass (right). In each panel, the mean values of random (open squares) and systematic (open triangles) errors are shown, along with their standard deviations by an error bar.\label{fig:ensemble_error}} \end{figure*}

Figure~\ref{fig:ensemble_error} shows error distributions of ensemble averages of RC mass from the five ZAHB models used in Figure~\ref{fig:ensemble}. The top panels show the mean errors and standard deviations from the original ZAHB models, while the bottom panels show error distributions using corrected sets of ZAHB models based on M67. Essentially, in all of the [Fe/H] and mass bins, the average random errors (blue open squares) exceed the sizes of the systematic errors (red open triangles) from the original models by a factor of $\ga2$. However, zero-point corrections lead to a better internal agreement of photometric masses from different ZAHB models and make systematic errors significantly smaller than random components. The dispersion in mass from the corrected models shows a mild increase toward higher metallicities and larger masses, while systematic errors from uncorrected models are consistently larger. Random errors from the corrected models are seemingly inflated compared to those from the original models, but this is due to the fact that valid solutions could not be found for a larger number of stars with uncorrected models.

\section{Summary and Discussion}\label{sec:discussion}

In this work, we derived masses of RC giants in the {\it Kepler} field by comparing the RC's positions on CMDs with theoretical ZAHB models. We utilized the fact that there is a relatively steep mass-luminosity relation of the ZAHB in the RC regime and constrained RC masses photometrically based on accurate distance and reddening estimates available for these stars. We took ZAHB models from various groups to judge the size of errors from theory. To reduce the impacts of potential systematic errors in the models, we adjusted the colors and magnitudes of the ZAHB models based on the observed location of RC giants in M67, assuming mild mass loss in this solar-metallicity cluster. Our working hypothesis on the mass loss in M67 effectively sets an upper limit on our photometric mass estimates.

One of our main findings is that the existing models are not consistent with each other, and even forcing agreement in M67 does not suppress metallicity trends. This has been examined by taking field RC giants with accurate distances from the {\it Gaia} and those in the metal-rich cluster NGC~6791. Our field star sample covers a wide range in metallicity ($-0.5\la{\rm [Fe/H]}\la+0.5$) and mass ($0.5\ M_\odot \la M_* \la 2\ M_\odot$) and therefore provides an opportunity to check the asteroseismic mass scale as a function of metallicity. We found that our photometric solutions at supersolar metallicities are strongly dependent on the adopted models even after the M67-based corrections, and that an ensemble average of mass becomes smaller than an asteroseismic mass. The departure at the high-metallicity end ([Fe/H]$\sim0.4$) is of a high significance ($>5\sigma$) due to a large number of metal-rich giants in the sample.

We also found similar trends in NGC~6791. Our photometric RC mass estimates range from $\sim0.7$ to $\sim1.1\ M_\odot$ from a number of model combinations, whereas the average mass ($0.88\pm0.16\ M_\odot$) is smaller than the asteroseismic mass \citep[$1.16\pm0.04\ M_\odot$;][]{pinsonneault:18}. The amount of mass loss on the RGB can be directly computed in this case, since the mass at the bottom of the RGB in NGC~6791 is sufficiently well known from the eclipsing binary study \citep[$1.15\pm0.02\ M_\odot$;][]{brogaard:11,brogaard:12}. Our photometric mass estimates yield a mean integrated amount of mass loss $\Delta M=0.27\pm0.16\ M_\odot$. Some of the models (Victoria and Y$^2$) predict moderately enhanced mass loss ($\sim0.4\ M_\odot$), in contrast to $\sim0.10\ M_\odot$--$0.14\ M_\odot$ expected from the Reimers formulation with $\eta \sim0.3$--$0.4$ as in the Galactic globular clusters. Nonetheless, the model-to-model dispersion is large, as shown by the large error in $\Delta M$. On the other hand, the asteroseismology suggests that the RC stars in NGC~6791 have lost a negligible fraction of their initial masses while ascending the RGB, suggesting a strongly bimodal mass-loss process (EHB versus RC) in the cluster. It may be that the asteroseismic masses are too close to the RGB mass, partly due to systematic errors in the $\Delta \nu$ and $\nu_{\rm max}$ measurements of RC giants \citep{pinsonneault:18}.

Our current understanding of stellar populations in the Galactic disk favors asteroseismic masses for the APOKASC-{\it Gaia} sample. The abundances of the APOKASC-{\it Gaia} sample mostly follow a low-$\alpha$ sequence in the [Fe/H] versus [$\alpha$/Fe] plane \citep[e.g.,][]{martig:15}, suggesting that most of them are associated with the traditional thin disk population. Proper motions and radial velocities of these stars are also consistent with thin disk kinematics on the Toomre diagram \citep[e.g.,][]{epstein:14}. The chemical and kinematical properties of a small fraction of the sample stars are consistent with the thick disk. Therefore, the bulk of the stars in the APOKASC-{\it Gaia} sample should have ages younger than $\sim4$--$8$~Gyr (with MSTO masses greater than $\sim1.15$--$1.3\ M_\odot$), and the masses of these RC giants are expected to be $\ga1\ M_\odot$ if a canonical mass-loss law on the RGB is assumed. In contrast, our photometric masses are too small for the ages of the stars and imply that most stars, especially at high metallicities, have experienced significantly enhanced mass loss ($\sim0.5\ M_\odot$), which is unlikely to be the case. Furthermore, theoretical or empirical corrections on the solar scaling relations relative to those of the RGB are of the order of a few percent, whereas the size of the corrections has to be enormous (several tens of percents) to match the lowest values of our photometric mass estimates. 

The strong model dependence of our photometric RC mass estimates may be a manifestation of underlying systematic errors in the models \citep[e.g.,][]{castellani:00}. Standard HB models are still subject to various theoretical uncertainties, most likely due to our limited knowledge of the helium enrichment parameter (\S~\ref{sec:models}) and mixing/convection near the outer region of a helium-burning core \citep[see][]{bressan:15}. For example, \citet{constantino:15} demonstrated that the asymptotic $l=1$ period spacing of gravity modes ($\Delta \Pi_1$) predicted from the standard HB models is systematically lower than observations, which can be partially relieved by additional core overshooting in the models \citep[see also][]{bossini:15}. Nevertheless, different modes of mixing in the core mostly increase the central helium-burning timescale, while the impact on the luminosity can be relatively mild \citep{constantino:15}. In any case, without additional constraints on model parameters and fine-tuning of core helium-burning models, it would be premature to draw a firm conclusion on the RGB mass loss.

\appendix

\section{Additional Model Comparisons}

\begin{figure}
\epsscale{1}
\plotone{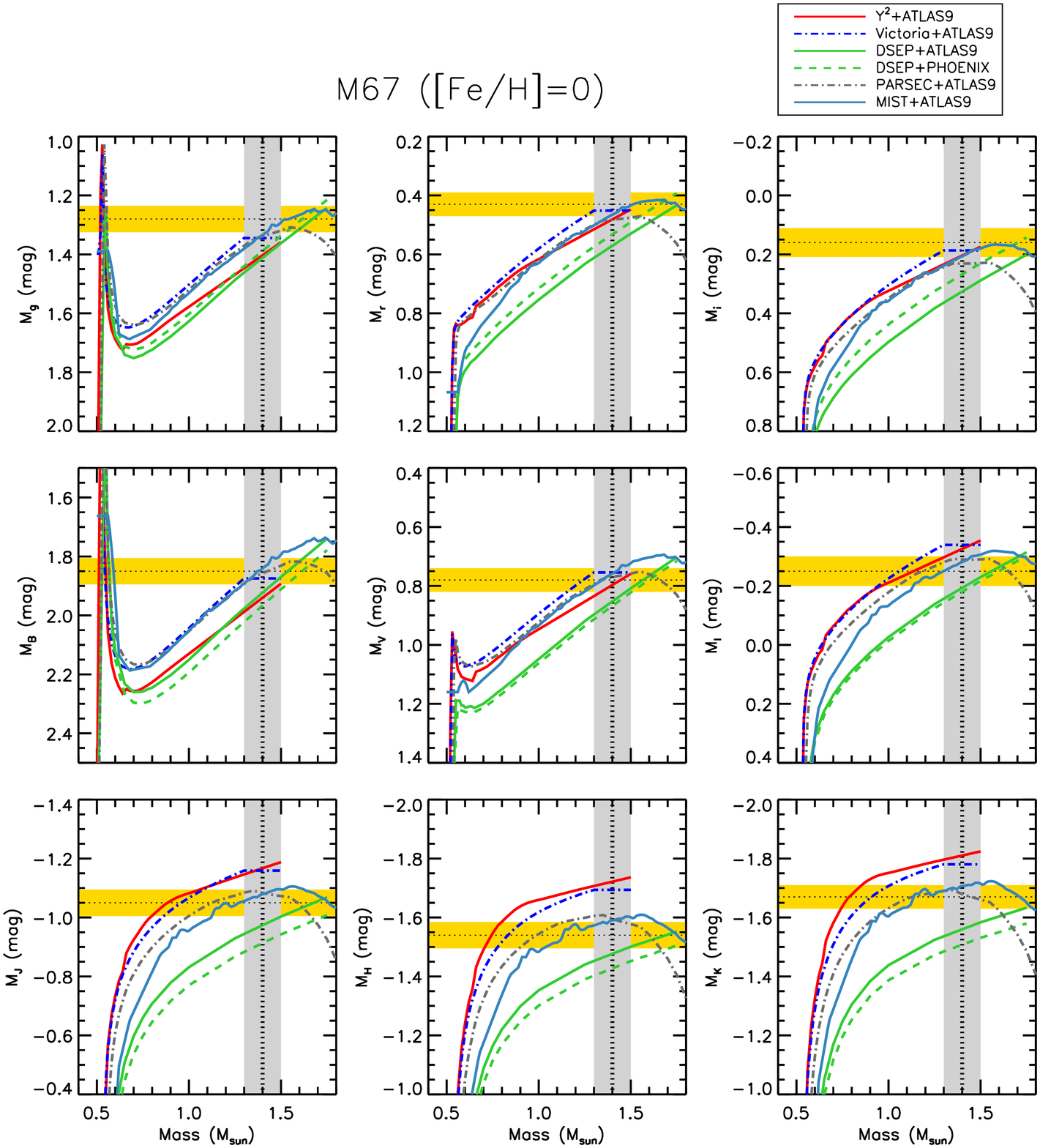}
\caption{Mass-luminosity relations of theoretical ZAHB models utilized in this work in various filter passbands. Models are shown at solar metallicity. The vertical gray bars indicate a mean asteroseismic mass of RC giants in M67 \citep{stello:16} and its $\pm1\sigma$ bound. The horizontal orange bars show a $\pm1\sigma$ range of the mean absolute magnitudes of the cluster's RC giants. Original models are shown without zero-point corrections on colors and magnitudes.\label{fig:model1}} \end{figure}

\begin{figure}
\epsscale{1}
\plotone{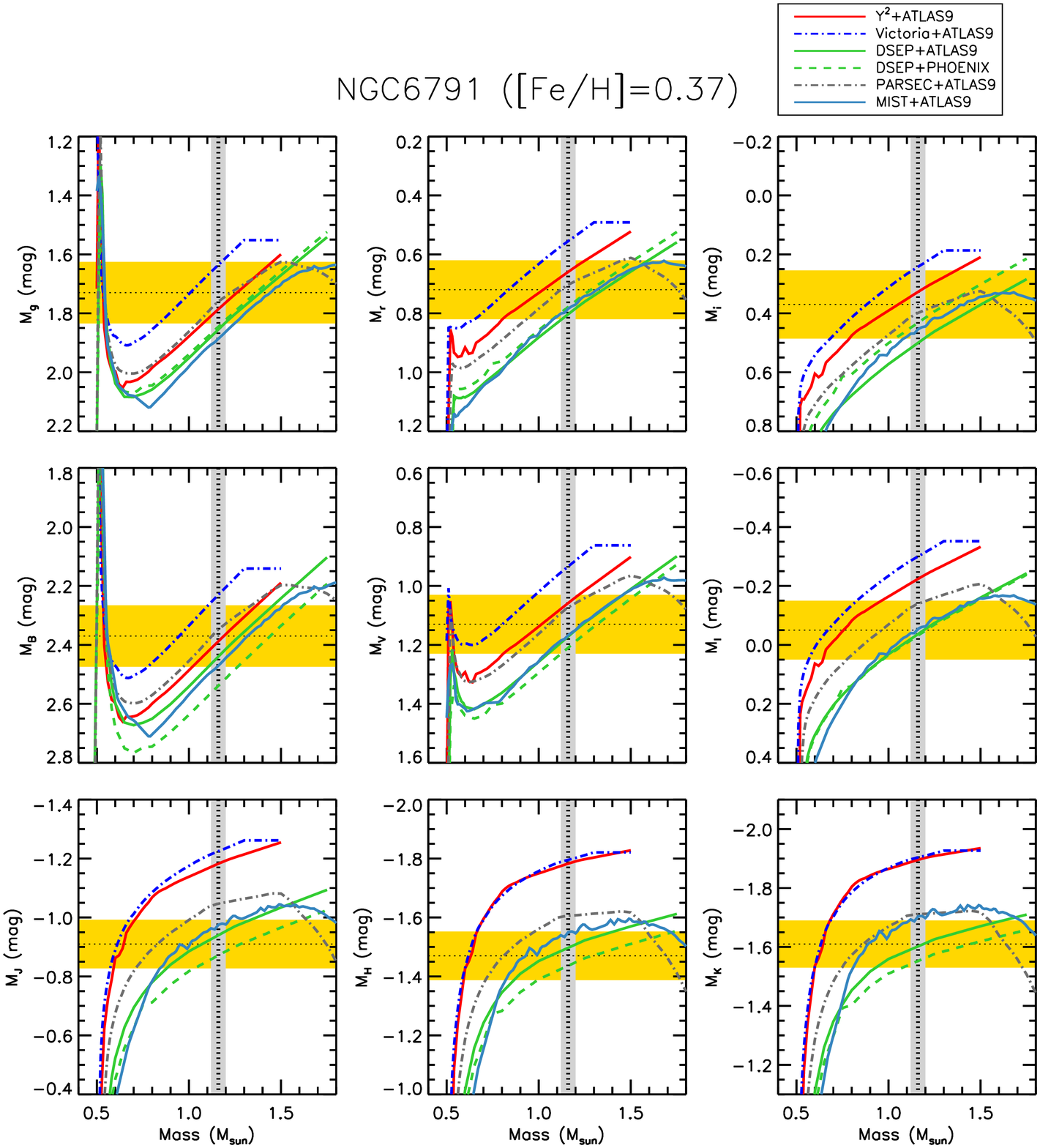}
\caption{Same as Figure~\ref{fig:model1} but for ZAHB models at [Fe/H]$=0.37$. The vertical gray and horizontal orange bars indicate the average and $\pm1\sigma$ measurement errors of the asteroseismic mass \citep{pinsonneault:18} and absolute magnitudes of RC giants in NGC~6791, respectively. Original models are shown without zero-point corrections on colors and magnitudes.\label{fig:model2}} \end{figure}

Figures~\ref{fig:model1} and \ref{fig:model2} show comparisons of mass-luminosity relations from theoretical ZAHB models utilized in this work in each of the $griBVI_CJHK_s$ filter passbands at solar and [Fe/H]=0.37, respectively. Original models are shown without zero-point corrections on colors and magnitudes. For comparison, the mean asteroseismic masses of RC giants in M67 and NGC~6791 \citep{stello:16,pinsonneault:18} are shown by vertical gray bars with $\pm1\sigma$ bounds. In addition, the mean absolute magnitudes of RC giants in each cluster (Table~\ref{tab:absmag}) are indicated by the horizontal orange bars.

\begin{figure}
\epsscale{0.45}
\plotone{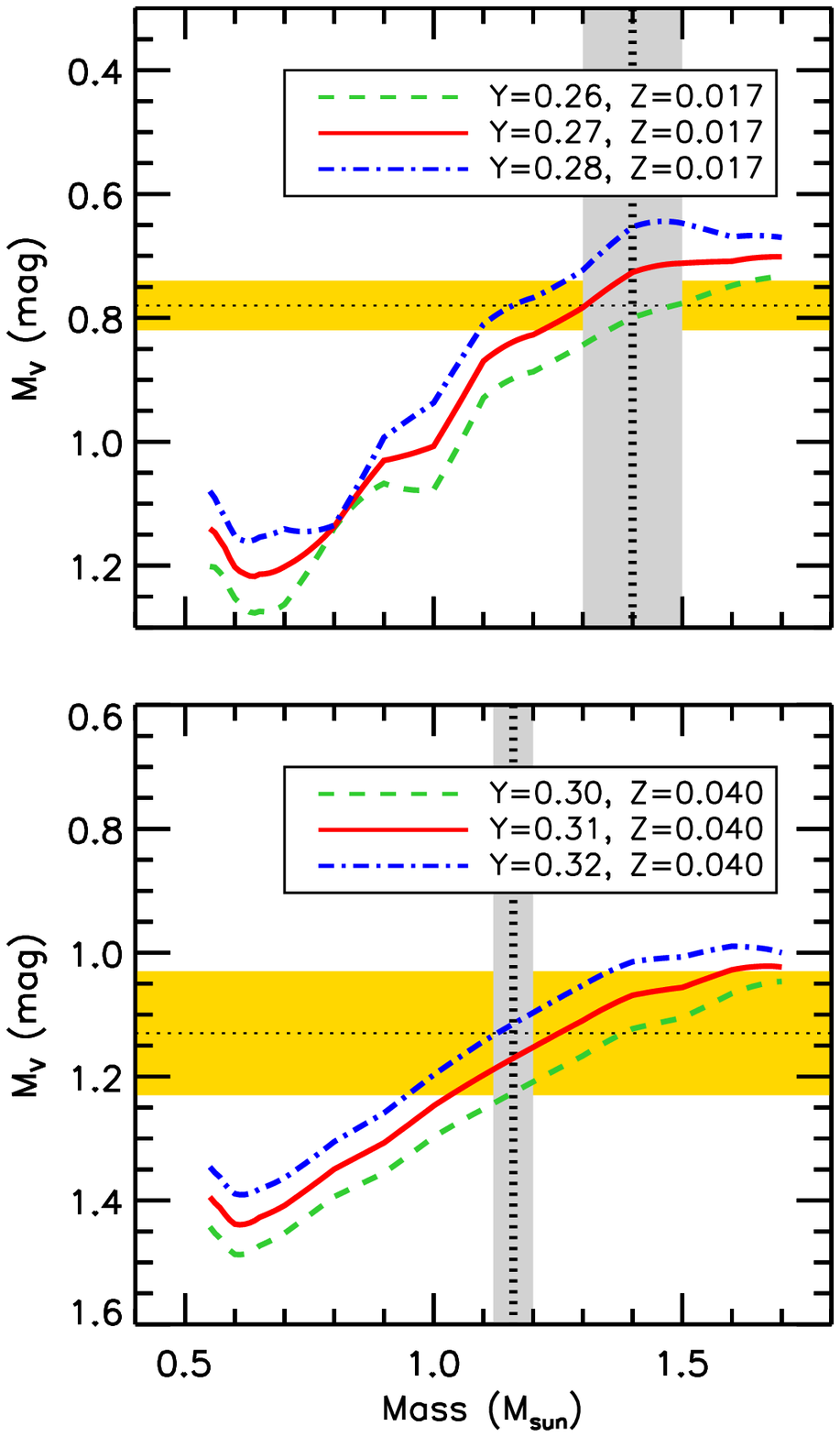}
\caption{Theoretical ZAHB models \citep{bertelli:08} with different initial helium mass fractions at $Z=0.017$ (top) and $Z=0.040$ (bottom). The vertical gray and horizontal orange bars indicate the average and $\pm1\sigma$ measurement errors of the asteroseismic mass and absolute $V$ magnitude of RC giants in M67 and NGC~6791, respectively.\label{fig:model3}} \end{figure}

In Figure~\ref{fig:model3}, theoretical ZAHB models \citep{bertelli:08} with different initial helium mass fractions ($\Delta Y=\pm0.01$) are compared with each other at a fixed $Z$. The top panel shows models at solar metallicity, and the bottom panel shows models at the metallicity close to that of NGC~6791. As shown in Figure~\ref{fig:model3}, a $\Delta Y=0.01$ change results in $\Delta V\sim0.05$~mag, almost independent of metallicity and stellar mass.

\begin{figure}
\epsscale{0.45}
\plotone{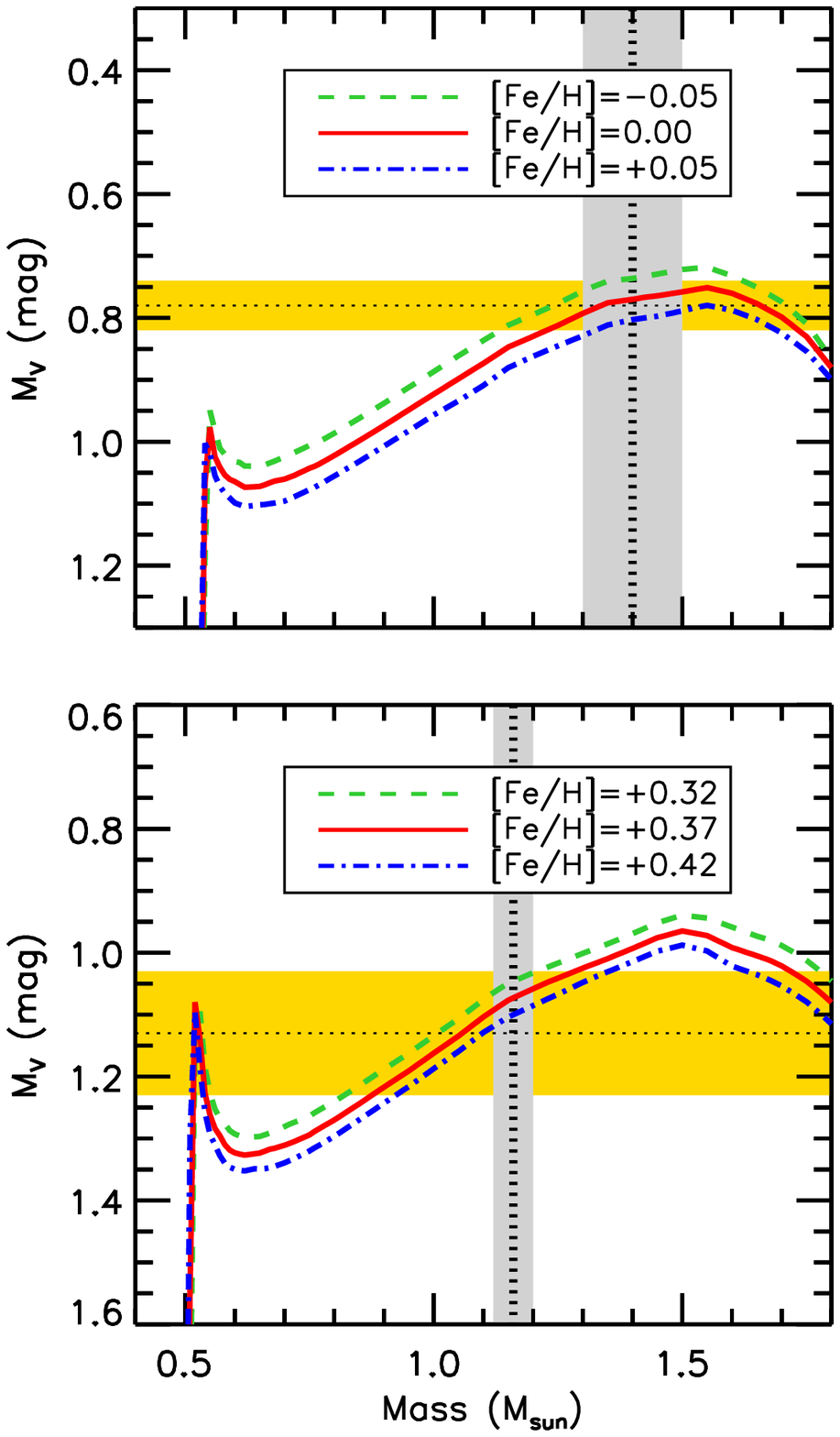}
\caption{Same as Figure~\ref{fig:model3} but showing the effect of metallicity from PARSEC ZAHB models.\label{fig:model4}} \end{figure}

In the top and bottom panels of Figure~\ref{fig:model4}, PARSEC ZAHB models are shown at [Fe/H]=0.0 and 0.37, respectively, along with those that differ by $\Delta{\rm [Fe/H]}=\pm0.05$. The $0.05$~dex change in metallicity results in an $\sim0.025$~mag change in $V$, which is nearly independent of metallicity and stellar mass, and is similar to those predicted by other models employed in this work.

\begin{figure}
\epsscale{2.05}
\plottwo{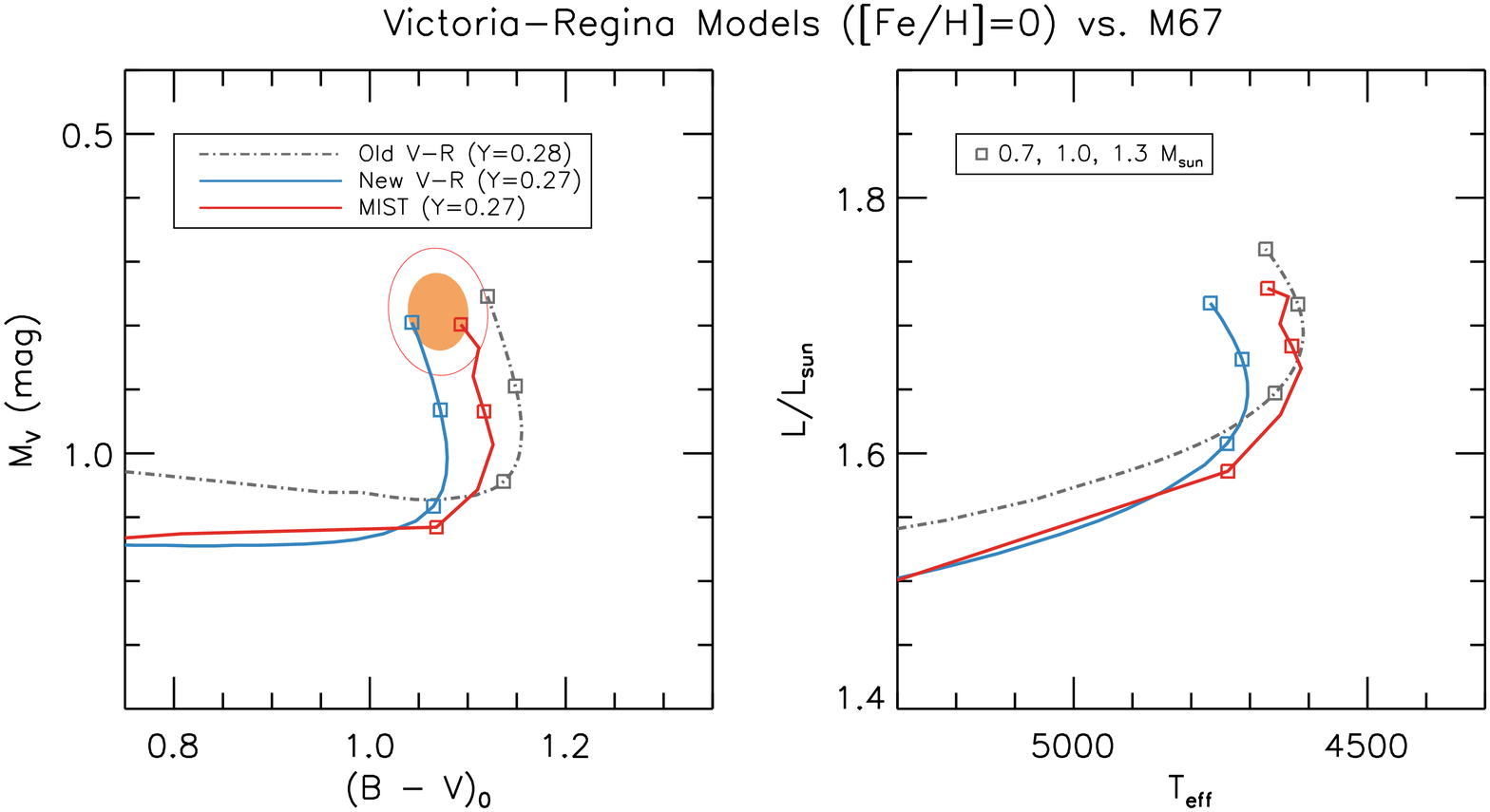}{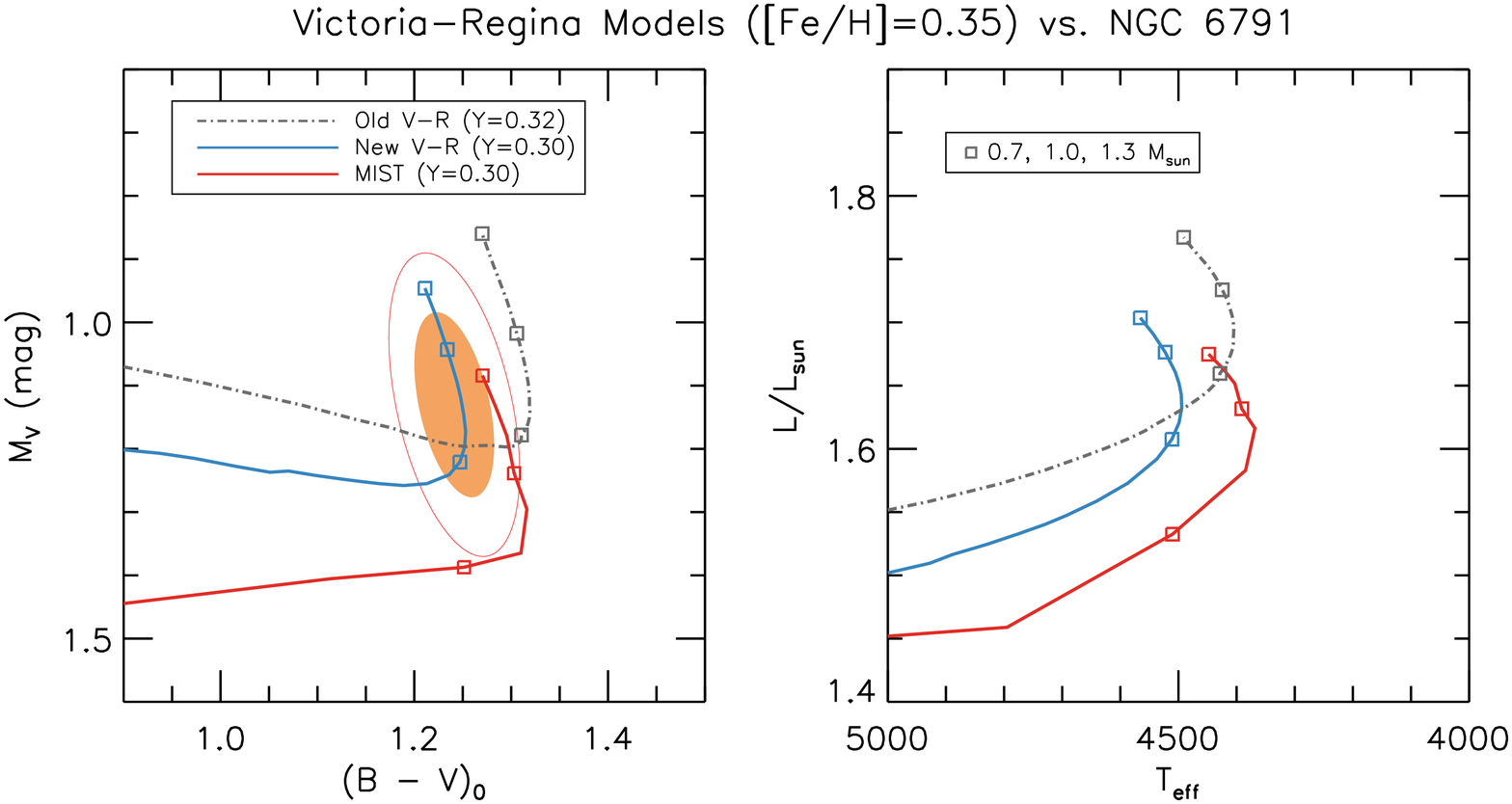}
\caption{Comparisons of the revised Victoria-Regina models (D.\ VandenBerg 2019, private communication) with the original models in \citet{vandenberg:06}. The former set of models was used in \citet{brogaard:12} in the estimation of RC mass in NGC 6791, and the latter set was employed in this work. For comparison, MIST ZAHB models are displayed by the red solid line. The contours indicate the 68\% and 95\% confidence intervals of the RC in M67 (top) and NGC~6791 (bottom), respectively (same as in Figures~\ref{fig:comp67} and \ref{fig:comp6791a}).\label{fig:model5}} \end{figure}

We employed the ZAHB models in \citet{vandenberg:06} in the main analysis of this work. However, major updates on these models have been made since its publication, including helium diffusion, and a subset of updated models was previously employed in \citet{brogaard:12} in the estimation of RC mass in NGC~6791. Figure~\ref{fig:model5} compares these models at solar metallicity (top panels) and [Fe/H]$=0.35$ (bottom panels), respectively. There are some systematic differences in the input parameters between the two model sets. The updated models are based on the solar abundance ratios in \citet{asplund:09}, while the original models are based on \citet{grevesse:98}. The difference in $V$ from different abundance mixtures is $\sim0.015$~mag at [Fe/H]=$0.35$ for a given stellar mass. In addition, the new models assume $Y=0.30$ from the eclipsing binary study \citep{brogaard:11}, while the old models have $Y=0.32$ from a steeper helium enrichment parameter (Table~\ref{tab:yz}). As shown in Figure~\ref{fig:model3}, $\Delta Y=0.02$ corresponds to $\Delta V\sim0.1$. There are also offsets expected from different bolometric corrections adopted by each set of models (ATLAS9 vs.\ MARCS), but the difference between them is negligible in $V$. The net effect is to lower the luminosity of the updated Victoria-Regina models by $\Delta V\sim0.05$ and subsequently increase a photometric mass by $\Delta M_*\sim0.1\ M_\odot$ from $V$ data only.

\acknowledgements

We thank the anonymous referee for providing detailed and helpful comments on the manuscript that eventually led us to include additional details on model comparisons in the Appendix. D.A.\ is greatly indebted to Donald VandenBerg for helpful discussions on Victoria-Regina models. We thank Jennifer Johnson, Leo Girardi, and Aaron Dotter for checking the manuscript and providing useful comments. D.A.\ and C.C.\ acknowledge support provided by the National Research Foundation (NRF) of Korea to the Center for Galaxy Evolution Research (No.\ 2017R1A5A1070354). D.A.\ acknowledges partial support by the NRF of Korea (NRF-2018R1D1A1A02085433). D.M.T.\ received partial support from National Science Foundation grant AST-1411685 to the Ohio State University.

Funding for the Sloan Digital Sky Survey IV has been provided by the Alfred P. Sloan Foundation, the U.S. Department of Energy Office of Science, and the Participating Institutions. The SDSS-IV acknowledges support and resources from the Center for High-Performance Computing at the University of Utah. The SDSS website is www.sdss.org.

SDSS-IV is managed by the Astrophysical Research Consortium for the  Participating Institutions of the SDSS Collaboration, including the Brazilian Participation Group, the Carnegie Institution for Science, Carnegie Mellon University, the Chilean Participation Group, the French Participation Group, Harvard-Smithsonian Center for Astrophysics, Instituto de Astrof\'isica de Canarias, Johns Hopkins University, Kavli Institute for the Physics and Mathematics of the Universe (IPMU) / University of Tokyo, Lawrence Berkeley National Laboratory, Leibniz Institut f\"ur Astrophysik Potsdam (AIP),  Max-Planck-Institut f\"ur Astronomie (MPIA Heidelberg), Max-Planck-Institut f\"ur Astrophysik (MPA Garching), Max-Planck-Institut f\"ur Extraterrestrische Physik (MPE), National Astronomical Observatories of China, New Mexico State University, New York University, the University of Notre Dame, Observat\'ario Nacional / MCTI, the Ohio State University, Pennsylvania State University, Shanghai Astronomical Observatory, the United Kingdom Participation Group, Universidad Nacional Aut\'onoma de M\'exico, the University of Arizona, the University of Colorado Boulder, the University of Oxford, the University of Portsmouth, the University of Utah, the University of Virginia, the University of Washington, the University of Wisconsin, Vanderbilt University, and Yale University.

This publication makes use of data products from the Two Micron All Sky Survey, which is a joint project of the University of Massachusetts and the Infrared Processing and Analysis Center/California Institute of Technology, funded by the National Aeronautics and Space Administration and the National Science Foundation.

This paper includes data collected by the {\it Kepler} mission. Funding for the {\it Kepler} mission is provided by the NASA Science Mission directorate.

This work has made use of data from the European Space Agency (ESA) mission {\it Gaia} (\url{https://www.cosmos.esa.int/gaia}), processed by the {\it Gaia} Data Processing and Analysis Consortium (DPAC; \url{https://www.cosmos.esa.int/web/gaia/dpac/consortium}). Funding for the DPAC has been provided by national institutions, in particular, the institutions participating in the {\it Gaia} Multilateral Agreement.

\end{document}